\documentclass[usenatbib]{mnras}
\usepackage{epsfig}
\usepackage{psfig}
\usepackage{amsmath,amssymb}
\usepackage{txfonts}
\usepackage{soul}
\allowdisplaybreaks
\def\av#1{\left\langle#1\right\rangle}

\newcommand\avrho{{\overline{\rho}}}
\newcommand\avxiM{\av{\xi}_{M}}
\newcommand\avxiN{\av{\xi}_{N}}

\renewcommand\d{{\rm d}}
\newcommand\Deltac{\Delta_{\rm c}}
\newcommand\Deltavir{\Delta_{\rm vir}}

\newcommand\ecrit{e_{\rm crit}}

\newcommand\Etwob{{E_{\rm 2b}}}

\newcommand\Eq{equation~}
\newcommand\Eqs{equations~}

\newcommand\fmajor{f_{\rm major}}

\newcommand\Fig{Fig.~}
\newcommand\Figs{Figs.~}

\newcommand\Lmax{{L_{\rm max}}}

\newcommand\M{M}

\newcommand\Mtwob{{\M}_{\rm 2b}}

\newcommand\Mcen{{\M}_{\rm cen}}
\newcommand\Maccr{{\M}_{\rm accr}}

\newcommand\Mvir{{\M}_{\rm vir}}
\newcommand\Mtwohc{{M_{200c}}}

\newcommand\Mfivetc{{M_{5000c}}}
\newcommand\Mtentc{{M_{10000c}}}
\newcommand\Mtwentytc{{M_{20000c}}}
\newcommand\MDelta{{\M}_{\Delta}}
\newcommand\MDeltasat{{\M}_{\rm \Delta,sat}}
\newcommand\MDeltahost{{\M}_{\rm \Delta,host}}

\newcommand\Mpc{{\rm \,Mpc}}
\newcommand\Msun{{\M}_{\odot}}

\newcommand\Omegam{\Omega_{\rm m}}
\newcommand\OmegaLambda{\Omega_\Lambda}

\newcommand\Psizero{\Psi_0}
\newcommand\rcen{r_{\rm cen}}

\newcommand\rs{r_{\rm s}}

\newcommand\rDelta{{r}_{\Delta}}
\newcommand\rDeltasat{{r}_{\rm \Delta,sat}}
\newcommand\rDeltahost{{r}_{\rm \Delta,host}}
\newcommand\rhocrit{\rho_{\rm crit}}

\newcommand\rstarhalf{r_{*,1/2}}
\newcommand\rvir{{r_{\rm vir}}}
\newcommand\rsnap{r_{\rm snap}}
\newcommand\rtwohc{r_{200c}}
\newcommand\rtwotc{r_{2000c}}
\newcommand\rfivetc{{r_{5000c}}}
\newcommand\rtentc{{r_{10000c}}}
\newcommand\rtwentytc{{r_{20000c}}}

\newcommand\rv{{\bf r}}

\newcommand\sigmaeight{\sigma_8}
\newcommand\Sect{Section~}
\newcommand\Sects{Sections~}

\newcommand\Tab{Table~}
\newcommand\Tabs{Tables~}

\newcommand\vsnap{v_{\rm snap}}

\newcommand\vr{v_r}

\newcommand\vcirc{v_{\rm circ}}
\newcommand\vtan{v_{\rm tan}}

\newcommand\vvsnap{{\vv_{\rm snap}}}
\newcommand\vtwob{v_{\rm 2b}}
\newcommand\vvtwob{\vv_{\rm 2b}}
\newcommand\vtantwob{v_{\rm tan,2b}}
\newcommand\vtansnap{v_{\rm tan,snap}}
\newcommand\vrtwob{v_{r,{\rm 2b}}}

\newcommand\vv{{\bf v}}

\newcommand\ximin{\xi_{\rm min}}
\newcommand\ximajor{\xi_{\rm major}}
\newcommand\ximed{\xi_{\rm med}}

\newcommand\zHF{z_{\rm HF}}
\defcitealias{Nip17}{N17}
\defcitealias{Jia15}{J15}

\begin{document}
%\begin{landscape}

\date{Resubmitted, 26 January 2018}

\title[Accretion onto central galaxies in clusters]{Accretion of satellites onto central galaxies in clusters: merger mass ratios and orbital parameters}

\author[C. Nipoti, C. Giocoli and G. Despali]{Carlo
  Nipoti$^{1}$\thanks{E-mail: carlo.nipoti@unibo.it}, Carlo Giocoli$^{1,2,3}$ and Giulia
  Despali$^{4}$
  \\
$^1$Department of Physics and Astronomy, Bologna University,
  via Gobetti 93/2, I-40129 Bologna, Italy \\
$^2$INAF - Osservatorio Astronomico di Bologna, via Gobetti 93/3, I-40129, Bologna, Italy \\ 
$^3$INFN - Sezione di Bologna, viale Berti Pichat 6/2, I-40127, Bologna, Italy \\  
$^4$Max Planck Institute for Astrophysics, Karl-Schwarzschild-Strasse 1, D-85740 Garching, Germany\\
}

\maketitle
\begin{abstract}
We study the statistical properties of mergers between central and
satellite galaxies in galaxy clusters in the redshift range $0<z<1$,
using a sample of dark-matter only cosmological $N$-body simulations
from Le SBARBINE dataset.  Using a spherical overdensity algorithm to
identify dark-matter haloes, we construct halo merger trees for
different values of the overdensity $\Deltac$. While the virial
overdensity definition allows us to probe the accretion of satellites
at the cluster virial radius $\rvir$, higher overdensities probe
satellite mergers in the central region of the cluster, down to
$\approx0.06\rvir$, which can be considered a proxy for the accretion
of satellite galaxies onto central galaxies.  We find that the
characteristic merger mass ratio increases for increasing values of
$\Deltac$: more than $60\%$ of the mass accreted by central galaxies
since $z\approx 1$ comes from major mergers. The orbits of satellites
accreting onto central galaxies tend to be more tangential and more
bound than orbits of haloes accreting at the virial radius. The
obtained distributions of merger mass ratios and orbital parameters
are useful to model the evolution of the high-mass end of the galaxy
scaling relations without resorting to hydrodynamic cosmological
simulations.
\end{abstract}

\section{Introduction}

Central Galaxies (CGs) in galaxy groups and clusters are typically
massive early-type galaxies with relatively old stellar populations
and little ongoing star formation.  CGs are believed to form in two
phases \citep{Mer85,Tre90,Dub98,Rus09,Lau14}. A first phase of {\em in
  situ} star formation at redshift $z\gtrsim 1$ is followed by a
second phase of growth via the so-called galactic cannibalism process
\citep{Ost75,Whi76,Hau78}, that is accretion of satellite galaxies
driven by dynamical friction \citep{Cha43}. Quantitatively, both
theoretical \citep{Del07,Fel10,Ton12,Sha15} and observational
\citep{Mar14,Bel16,Buc16,Vul16b} arguments suggest that about half of
the stellar mass of CGs is assembled {\em in situ} at $z\gtrsim 1$,
and the other half is assembled at relatively late times ($z\lesssim
1$) via cannibalism processes.

The effect of this cannibalism-driven growth phase on the properties
of the CG (for instance size and velocity dispersion) is determined
not only by the properties of the cannibalised galaxies
\citep[e.g.\ mass ratio between satellite and central;][]{Naa09}, but
also by the merging orbital parameters \citep{Boy06,Nip12}.  Measures
of size, velocity dispersion, luminosity and stellar mass of observed
CGs lie on tight scaling relations \citep{Ber07,Liu08,Vul14}. Knowing
the properties of the mergers that occur during the late growth of CGs
is thus important to theoretically understand the origin and evolution
of their scaling relations.

\citet[][hereafter \citetalias{Nip17}]{Nip17} made the point that,
given the very special location of CGs, at the bottom of the deep
potential well of the host group or cluster, the distribution of the
orbital parameters of the central-satellite encounters can be quite
different from that of the encounters between galaxies not belonging
to groups or clusters.  \citetalias{Nip17} has characterised the
distribution of the orbital parameters for central-satellite mergers
using idealised $N$-body simulations in which the host system (a
cluster or a group) is modelled as an isolated, spherical,
collisionless $N$-body system and the satellite is rigid, being
represented by a single massive, softened particle. In particular, the
simulations of \citetalias{Nip17} are not framed within a cosmological
context: the initial orbital parameters of the satellites are
extracted from the host-halo distribution function, based on the
assumption that violent relaxation \citep{Lyn67} is rapid and the
satellite population does not retain much memory of the cosmological
distribution of the orbital parameters at time of infall. The orbital
parameters of the satellites then evolve due to dynamical friction
(i.e.\ the satellites lose orbital energy and angular momentum).

In this work we improve on the analysis of \citetalias{Nip17} by
considering the problem in a fully cosmological setting, focusing on
the growth of CGs in clusters of galaxies. For this purpose, we take
advantage of the suite of cosmological simulations Le SBARBINE
\citep{Des16}.  These simulations are dark-matter only and so they do
not contain a realistic galaxy population. Nevertheless, if we assume
that CGs sit at the center of dark-matter haloes and if we select the
central regions of these host haloes at overdensities typical for the
location of the central galaxies, they can be used for our purposes.
Following \citet{Des16}, we identified haloes in Le SBARBINE
simulations for different overdensity threshold: $\Deltac=\Deltavir$,
where $\Deltavir$ is the redshift-dependent virial overdensity, and
$\Deltac=200$, $5000$, $10000$ and $20000$, independent of
redshift. For each of these halo definitions, we also built the
corresponding merger history trees.  When $\Deltavir$ is considered,
the entire virialised region of the halo is selected (in the case of a
galaxy cluster, the entire cluster-size dark-matter halo). When higher
overdensities are considered, smaller regions of the halo are
selected: for sufficiently high overdensity we select only the central
part of the virialised halo, which we can roughly identify with the
central galaxy. Moreover at these high overdensities, dense
substructures within the main virial halo can be identified as
independent structures.  For each given overdensity we measure the
properties of the mergers (specifically, the mass ratios and the
orbital parameters). We then compare the properties of cosmological
accretion (at the virial radius $\rvir$) with those of accretion onto
central galaxies (at some inner radius $r\ll\rvir$).

The paper is organised as follows. \Sect\ref{sec:numerical} describes
the simulations and the numerical methods. The orbital parameters of
halo-halo encounters are defined in \Sect\ref{sec:orbital_parameters}.
The results are presented in \Sect\ref{sec:results}, while in
\Sect\ref{sec:conclusions} we draw our conclusions.

\section{Numerical methods}
\label{sec:numerical}
%\subsection{Main properties of the $N$-body simulations}
\subsection{$N$-body simulations and identification of dark-matter haloes}
\label{sec:simu}

In this work we make use of the results from the cosmological
dark-matter only $N$-body simulations Le SBARBINE \citep{Des16}.  The
assumed background cosmology and initial conditions of the simulations
are consistent with the results from the Planck Collaboration XVI
\citep{Pla14}. In particular, in the simulations and throughout this
paper we adopt a standard $\Lambda$ cold dark matter cosmological
model with the following parameters: matter density parameter $\Omegam
= 0.307$, cosmological constant density parameter $\OmegaLambda=
0.693$, linear power spectrum amplitude $\sigmaeight = 0.829$ and
dimensionless Hubble constant $h=0.677$.  Here we use only the two
highest resolution runs among Le SBARBINE simulations: Ada
(dark-matter particle mass $m=2.87\times 10^7\Msun $) and Bice
($m=2.29\times 10^8\Msun$).  The two simulations have the same number
of particles ($N=1024^3$), but different box size: $92.3\Mpc$ for Ada
and $184.6\Mpc$ for Bice.

%$62.5h^{-1}\Mpc$ for Ada and $125h^{-1}\Mpc$ for Bice.

At each stored snapshot haloes are identified using a
  spherical overdensity algorithm (e.g.
  \citealt{tormen98,tormen04,giocoli08}).  In practice, haloes are
defined as spherical overdensities with radius such that the average
density is
\begin{equation}
\avrho=\Deltac\rhocrit,
\end{equation}
where $\Deltac$ is the critical overdensity and
\begin{equation} 
\rhocrit(z)= \frac{3H^2(z)}{8\pi G} 
\label{eq:rhocrit}
\end{equation}
is the critical density of the Universe, depending on redshift through
the Hubble parameter $H(z)$.  Using this method, the overdensity
threshold that defines the halo boundaries can be varied depending on
the observational data that one wants to compare with. While the
virial overdensity $\Deltac=\Deltavir$ is commonly used in structure
formation studies, other definitions can be chosen to be more similar
to observational data sets: $\Deltac=500$ is typically used in X-ray
observations to define the mass of a galaxy cluster, while
$\Deltac=200$ is often used to fit weak lensing shear profiles. In our
case $\Deltac=20000$ corresponds to the region within $r\simeq
0.06r_{vir}$, which is a proxy for the size of the CG in a cluster.

In this work, we consider different choices for $\Deltac$. For each
value of $\Deltac$, we define the halo mass $\MDelta$ and the halo
radius $\rDelta$.  First of all, we consider the virial value
$\Deltac=\Deltavir(z)$, which depends on $z$ and on the cosmological
parameters, as given by \citet{Eke96}. For the assumed cosmology
$\Deltavir$ increases with redshift: reference values are
$\Deltavir\simeq 97.9$ at $z=0$ and $\Deltavir\simeq 154$ at $z=1$.
When $\Deltavir=\Deltac$ the halo radius and mass are, respectively,
the virial radius $\rvir$ and the virial mass $\Mvir$.  For comparison
with previous work we consider also the standard value $\Deltac=200$,
independent of redshift: in this case the halo mass and radius are
$\Mtwohc$ and $\rtwohc$, respectively.  Finally, in order to study the
behaviour of mergers in central parts of the haloes, we explore the
following other values of the critical overdensity, independent of
redshift: $\Deltac=5000$ (with mass $\Mfivetc$ and radius $\rfivetc$),
$\Deltac=10000$ (with mass $\Mtentc$ and radius $\rtentc$) and
$\Deltac=20000$ (with mass $\Mtwentytc$ and radius $\rtwentytc$).  It
must be noted that in each catalogue (i.e.\ for each value of the
considered overdensity $\Deltac$) the haloes are identified
independently. Therefore the number of haloes is in general different
in each catalogue, because a halo that, at a given redshift, is
identified for a given value of $\Deltac$, at the same redshift might
be ``incorporated'' in a bigger halo when a lower overdensity is
considered \citep[for more details see figure 2 of ][]{Des16}.

\subsection{Sample of haloes and catalogues of halo-halo encounters}
\label{sec:sample}

We study the redshift range $0\lesssim z\lesssim 1$, in which Le
SBARBINE simulations have 13 snapshots at the following redshifts:
$z_1=1.012$, $z_2=0.904$, $z_3=0.796$, $z_4=0.694$, $z_5=0.597$,
$z_6=0.507$, $z_7=0.421$, $z_8=0.34$, $z_9=0.264$, $z_{10}=0.192$,
$z_{11}=0.124$, $z_{12}=0.06$, $z_{13}=0$.  We create a sample of
galaxy clusters by selecting, in the $\Deltavir$ catalogue, all haloes
with $\Mvir\geq 10^{14}\Msun$ at $z=0$. The resulting sample consists
of 101 haloes at $z=0$ (12 haloes in Ada and 89 haloes in Bice).  We
identify these 101 haloes also in the higher-overdensity
($\Deltac>\Deltavir$) catalogues ($\Deltac=200$, $5000$, $10000$ and
$20000$), finding that all of them have counterparts in all the
considered catalogues. We note that, as the selection in mass is done
on $\Mvir$, the selected $z=0$ haloes can have mass
$\MDelta<10^{14}\Msun$ for $\Deltac>\Deltavir$, because the mass of a
given halo decreases for increasing $\Deltac$.

In order to identify {\em halo-halo encounters} we proceed as follows.
From the halo catalogues built for the 13 simulation snapshots and for
each overdensity, we construct the halo merging history tree.
Starting from each halo at $z = 0$, we define its progenitors at the
previous output, $z = 0.06$, as all haloes that, in the time elapsed
between two snapshots, have given at least 50\% of their particles to
the considered $z=0$ halo.  The \emph{main progenitor} at $z = 0.06$
is defined as the most massive progenitor of the $z=0$ halo.  We then
repeat the same procedure, now starting from the main progenitor at $z
= 0.06 $ and considering its progenitors at $z = 0.124$, and we
proceed backwards in time, always following the main progenitor halo.
The resulting merger tree consists of a main trunk, which traces the
main progenitor back in time, and of {\em satellites}; these last are
all the progenitors that at any time merge directly onto the main
progenitor. By construction, for given simulation and descendant halo,
the definition of the main progenitor depends on the time sampling
(i.e. the number of snapshots): in principle it is possible that the
main branch is not identified correctly if the time sampling is
insufficient\footnote{For instance, consider a halo with mass $M$ at
  redshift $z_i$ that at redshift $z_{i-1}>z_i$ splits in two haloes
  of mass $M_1$ and $M_2<M_1$, which, in turn, at redshift
  $z_{i-2}>z_{i-1}$ split, respectively, in two haloes of mass
  $M_{1,1}$ and $M_{1,2}<M_{1,1}$, and in two haloes of mass $M_{2,1}$
  and $M_{2,2}<M_{2,1}$. When $M_{2,1}>M_{1,1}$ the main branch is
  misidentified if the $z_{i-1}$ snapshot is missing. }.  However,
\citet{giocoli08thesis}, using simulations with time sampling similar
to Le SBARBINE, has shown that the probability of misidentifying the
main progenitor branch for galaxy cluster scale haloes is much below
10\% in the redshift interval $0\lesssim z\lesssim 1$. In addition the
consistence of our results with those obtained by \citet[][hereafter
  \citetalias{Jia15}]{Jia15} with the DOVE simulation
(Appendix~\ref{sec:orb_par_at_r200}), which has a better time
sampling, ensures us that the time sampling of our runs is good enough
to uniquely follow the main halo progenitor branch back in time.

We define the halo-halo encounters between two subsequent snapshots by
selecting in the higher-redshift snapshot all the pairs satellite-main
progenitor.  We indicate the physical properties of the main
progenitor (sometimes referred to also as host halo) with the
subscript ``host'' and those of the less massive progenitors
(satellites) with the subscript ``sat''.  For each pair we then
measure the masses ($\MDeltahost$ and $\MDeltasat$), radii
($\rDeltahost$ and $\rDeltasat$), relative position $\rv$ and relative
velocity $\vv$ of the centres of mass.  We apply the procedure
described above to all the $z=0$ haloes of our sample and to all their
main progenitors, back to the $z=0.904$ snapshot. In this way we build
our catalogue containing all the halo-halo encounters in the redshift
range $0<z<1$ that end up in cluster-size haloes at $z=0$. Of course
the number of encounters and their properties are different for
different overdensities.

We note that not all the halo-halo encounters must be considered rapid
mergers. By construction, our catalogue of halo-halo encounters
include cases of satellite haloes that, after the encounter, escape
from the main halo. In these cases the haloes are distinct at a given
snapshot, are identified as a single halo at a later snapshot, but are
again distinct at an even later time step. In the terminology of this
work we then distinguish halo-halo {\em encounters} and halo-halo {\em
  mergers}.  In \Sect\ref{sec:mergers_flybys} we will define a
criterion to select the subsample of mergers in the whole sample of
halo-halo encounters.

For each halo-halo encounter we define the mass ratio
$\xi=\MDeltasat/\MDeltahost$.  In order to have a robust measure of
the properties of halo-halo encounters and mergers, we limit our
exploration to $\xi\geq\ximin$, where $\ximin$ is a minimum mass ratio
such that the number of particles of the satellite is at least
$N\approx$ 100.  Clearly $\ximin$ depends on the mass resolution of
the simulations, on the explored redshift range and on the considered
overdensity $\Deltac$.  In particular, we adopt $\ximin=0.005$ for
$\Deltac=\Deltavir$ and $\Deltac=200$, $\ximin=0.01$ for
$\Deltac=5000$ and $\Deltac=10000$, and $\ximin=0.1$ for
$\Deltac=20000$.  We verified that with these choices, for our sample
of encounters in the redshift range $0<z<1$, the number of satellites
with $N<100$ never exceeds $5\%$ of the entire satellite population.

Using dark-matter only simulations to infer the properties of
central-satellite galaxy mergers in clusters, we are implicitly
assuming that the dynamical evolution of subhaloes orbiting massive
host haloes is not significantly influenced by the presence of
baryons. In fact, comparisons between dark-matter only and
hydrodynamic cosmological simulations indicate that some properties of
the subhalo mass function can be modified by the presence of baryons
\citep[][and references therein]{Chu17}. However, the effect is
negligible for the relatively high satellite-host mass ratios
considered in this work (see figure 1 of \citealt{Chu17}).

\section{Orbital parameters of halo-halo encounters}
\label{sec:orbital_parameters}

It is useful to describe an encounter between two haloes in terms of
the orbital parameters calculated in the point-mass two-body
approximation. This description, though not rigorous for extended
objects, is often used in the study and classification of mergers of
galaxies \citep[e.g.][]{Boy06,Nip09} and dark-matter haloes
\citep[e.g.][]{Kho06,Pos14}.  Here we define the point-mass two-body
approximation orbital parameters of halo-halo encounters, following
the formalism of \citetalias{Nip17}. The orbit can be fully
characterised by the pair of parameters orbital energy and angular
momentum.  For a halo-halo encounter we define the two-body
approximation orbital energy per unit mass
\begin{equation}
\Etwob=\frac{1}{2}v^2-\frac{G\Mtwob}{r},
\end{equation}
where $\Mtwob\equiv \MDeltahost+\MDeltasat$, $r\equiv|\rv|$ is the
relative distance and $v\equiv |\vv|$ is the relative speed, between
their centres of mass. It is useful to decompose $\vv$ in its radial
component $\vr=\vv\cdot\rv/r$ and its tangential component, with
modulus $\vtan=\sqrt{v^2-\vr^2}$.  The modulus of the orbital angular
momentum per unit mass is $L=r\vtan$.  At fixed separation $r$ and
energy $\Etwob$, the modulus of the maximum allowed specific angular
momentum is
\begin{equation}
\Lmax=r\sqrt{2\left(\Etwob+\frac{G\Mtwob}{r}\right)}=rv,
\label{eq:lmax}
\end{equation}
where $v$ is the relative speed when the two haloes have separation
$r$.  Clearly, $\Lmax$ is such that $L/\Lmax=\vtan/v$.

Another set of orbital parameters used to classify halo-halo
encounters (e.g.\ \citealt{Ben05}; \citetalias{Jia15}) is the pair
($v/\vcirc$, $|\vr|/v$), where, given a distance $r$ from the centre
of the host system (for instance the overdensity radius
$\rDeltahost$), $v$ is the relative speed at $r=\rDeltahost$, $\vr$ is
the radial component of the relative velocity at $r=\rDeltahost$ and
\begin{equation}
  \vcirc=\sqrt{\frac{G\MDeltahost}{\rDeltahost}}
\end{equation}
is the host circular velocity at $\rDeltahost$.

Given the finite time sampling (i.e.\ the finite number of snapshots),
in our simulations we have information on the halo-halo relative
velocity $\vvsnap$ when the two haloes have a separation $\rsnap$ that
is in general larger than $\rDeltahost$. As discussed in several
previous works (e.g.\ \citealt{Ben05}, \citetalias{Jia15}), it is thus
necessary to apply a correction to recover ($v/\vcirc$, $|\vr|/v$)
measured when the satellite crosses the desired overdensity radius of
the host (i.e.\ when the separation is $\rDeltahost$). We correct the
velocity as follows. We first compute the relative velocity $\vvtwob$
at $\rDeltahost$, assuming that the point-mass two-body energy and
angular momentum are conserved: $\vvtwob$ is such that
\begin{equation}
  \frac{1}{2}\vsnap^2-\frac{G(\MDeltahost+\MDeltasat)}{\rsnap}
  =\frac{1}{2}\vtwob^2-\frac{G(\MDeltahost+\MDeltasat)}{\rDeltahost}
\label{eq:en_cons}
\end{equation}
and 
\begin{equation}
  \rsnap\vtansnap=\rDeltahost\vtantwob,
  \label{eq:ang_mom_cons}
\end{equation}
where $\vtantwob$ is the tangential component of $\vvtwob$.  If, as in
most cases, \Eqs(\ref {eq:en_cons}) and (\ref{eq:ang_mom_cons}) give
$\vtantwob\leq \vtwob$, the modulus of the radial component of
$\vtwob$ is $|\vrtwob|=\sqrt{\vtwob^2-\vtantwob^2}$.  If, instead,
\Eqs(\ref {eq:en_cons}) and (\ref{eq:ang_mom_cons}) give
$\vtantwob>\vtwob$ (which indicates that the point-mass two-body orbit
is too crude an approximation), we simply fix $\vtantwob=\vtwob$ and
$\vrtwob=0$. Finally, we define the corrected velocity $\vv$ at the
time of crossing ($r=\rDeltahost$) to be such that
\begin{equation}
\left(\frac{v}{\vcirc}\right)_{\rDeltahost}=\frac{1}{2}\frac{\vsnap+\vtwob}{\vcirc}
\label{eq:vcorr}
\end{equation}
and 
\begin{equation}
\left(\frac{\vr}{v}\right)_{\rDeltahost}=\frac{1}{2}\left[\left(\frac{\vr}{v}\right)_{\rm snap}+\left(\frac{\vr}{v}\right)_{\rm 2b}\right].
\label{eq:vrcorr}
\end{equation}
We verified that this is a reasonably good approximation by comparing
the distributions of our sample of haloes with previous literature
work (see Appendix~\ref{sec:orb_par_at_r200}).

\section{Results}
\label{sec:results}

\subsection{Halo masses and radii at different overdensities}
\label{sec:masses_radii}

The ratios $\MDelta/\Mvir$ and $\rDelta/\rvir$ are decreasing
functions of $\Deltac$ \citep[see][]{Des17}. The exact values of these
ratios depend on the halo mass density distribution through the halo
concentration parameter \citep{giocoli12b}.  We computed
$\MDelta/\Mvir$ and $\rDelta/\rvir$ for all haloes in our sample
(i.e.\ the 101 $z=0$ haloes and all their main progenitors in all
previous snapshot back to $z=0.904$ (altogether 1224 haloes;
\Sect\ref{sec:sample}). The means and standard deviations of the
distributions of $\MDelta/\Mvir$ and $\rDelta/\rvir$ are reported in
\Tab\ref{tab:ratios} for $\Deltac=200$, $5000$, $10000$ and $20000$.
For the highest overdensity here considered $\Deltac=20000$, the
average values are $\MDelta/\Mvir\approx0.05$ and
$\rDelta/\rvir\approx0.06$.  The distributions of $\MDelta/\Mvir$ and
$\rDelta/\rvir$ are broader for increasing $\Deltac$, with standard
deviations in the range $10-17\%$ for $\rDelta/\rvir$ and $11-44\%$
for $\MDelta/\Mvir$.

In the following sections we will compare our results with those of
simulations of satellites in isolated host haloes
(\citetalias{Nip17}), in which the merger orbital parameters were
measured at a radius $\rcen=0.12\rs$, where $\rs$ is the halo scale
radius. If we identify the truncation radius of \citetalias{Nip17}
with $\rvir$, we have $\rcen \simeq 0.024\rvir$ and that the mass
contained within $\rcen$ is $\Mcen\simeq0.0075\Mvir$. Therefore the
region probed by \citetalias{Nip17} is somewhat smaller than the most
central region here considered ($\Deltac=20000$) and roughly
corresponds to an overdensity $\Deltac=50000$.  It is useful to note
that \citet{Kra13} finds that, on average, the three-dimensional
half-mass radius of the stellar distribution of observed central
galaxies is $\rstarhalf\approx0.015\rtwohc\approx0.012\rvir$ (using the
average ratio $\rtwohc/\rvir\simeq 0.85$ found for our sample of
haloes; see \Tab\ref{tab:ratios}). Therefore, in terms of $\rstarhalf$, we
have $\rtwentytc\approx 5 \rstarhalf$ and $\rcen\approx 2 \rstarhalf$, which
indicates that both $\rtwentytc$ and $\rcen$ probe the region of the
halo occupied by the stellar distribution of the central galaxy.

\begin{table}
%\begin{center}
\caption{Mean ($\mu$) and standard deviation ($\sigma$) of the
  distributions of $\MDelta/\Mvir$ and $\rDelta/\rvir$, for different
  values of $\Deltac$, for our sample consisting of the 101 $z=0$
  haloes with $\Mvir\geq 10^{14}\Msun$ and all their main progenitor
  haloes in the previous snapshots, back to $z=0.904$ (altogether 1224
  haloes).
 \label{tab:ratios}}
\begin{tabular}{rcccc}
\hline 
$\Deltac$  & $\mu(\MDelta/\Mvir)$ & $\sigma(\MDelta/\Mvir)$ & $\mu(\rDelta/\rvir)$ & $\sigma(\rDelta/\rvir)$  \\
\hline
200  & 0.848  & 0.097 &  0.807 & 0.082\\
5000  & 0.148  & 0.047 &  0.153 & 0.020\\
10000  & 0.085  & 0.032 &  0.100 & 0.015\\
20000  & 0.045  & 0.020 &  0.064 & 0.011\\
\hline
\end{tabular}
%\end{center}
\end{table}

\subsection{Mergers and fly-bys}
\label{sec:mergers_flybys}

%%%%%%%%%%%%%%FIG 1
\begin{figure}
  \centerline{\psfig{file=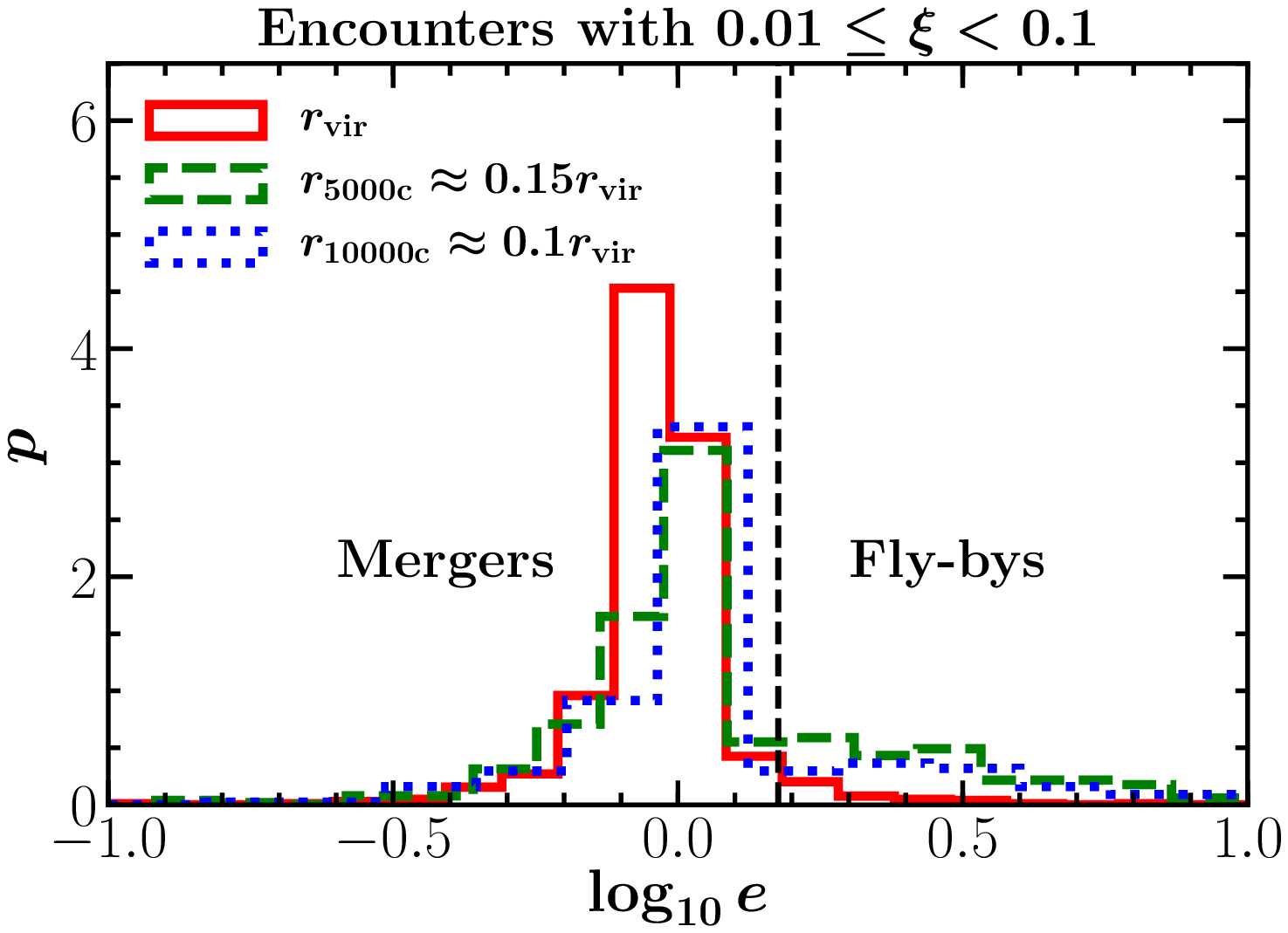,width=\hsize}}
 \centerline{\psfig{file=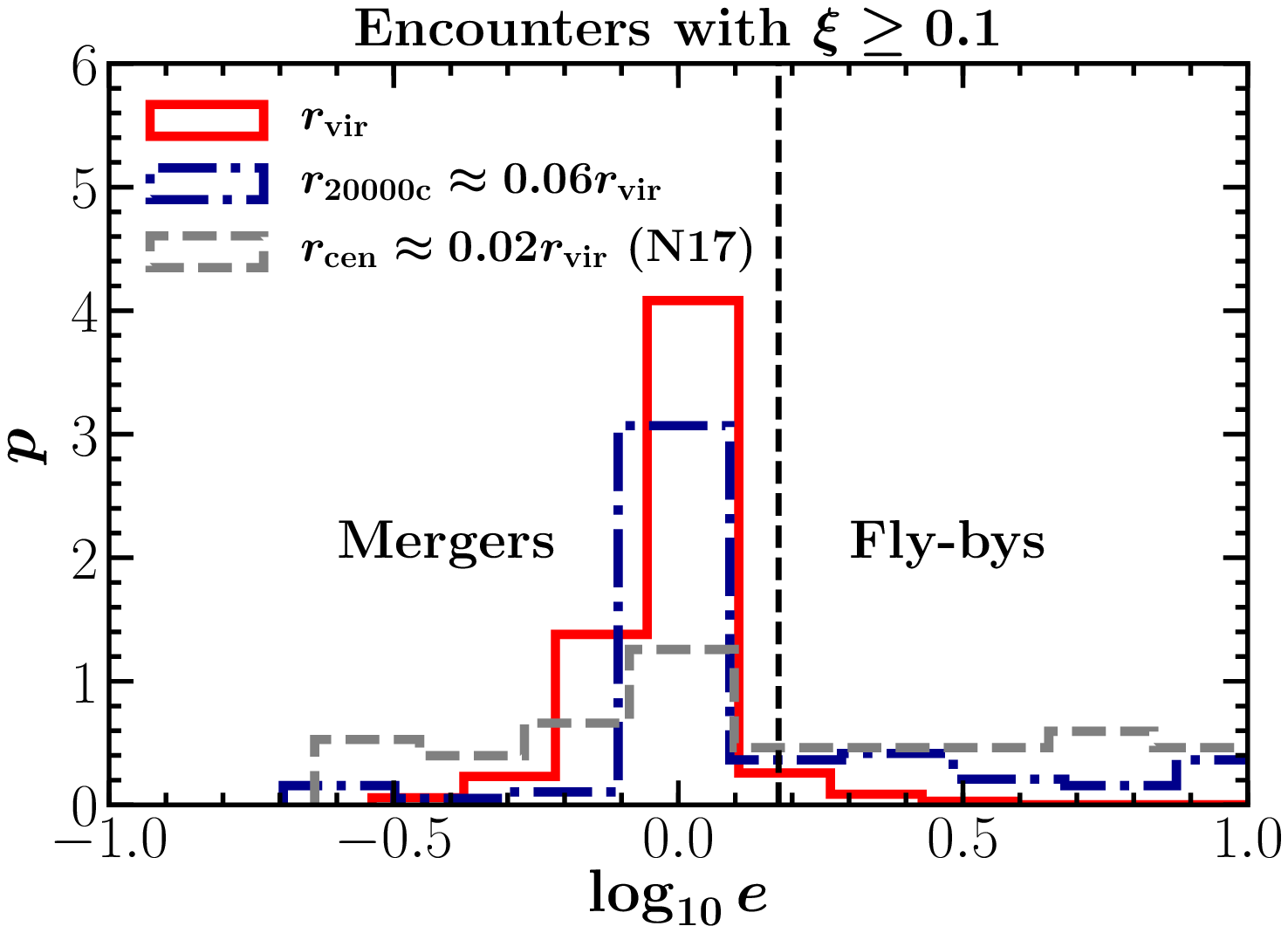,width=\hsize}}
\caption{{\em Upper panel}. Probability distribution $p=\d n/\d x$ of
  the logarithm of the orbital eccentricity of halo-halo encounters
  computed in the two-body approximation ($x=\log_{10}e$), for
  critical overdensities $\Deltac=\Deltavir$ (solid histogram),
  $\Deltac=5000$ (dotted histogram) and $\Deltac=10000$ (dashed
  histogram).  Here we consider mergers with mass ratios $0.01\leq \xi<
  0.1$. The vertical dashed line ($e=1.5$) discriminates mergers ($e
  \leq 1.5$) and fly-bys ($e>1.5$). {\em Lower panel}. Same as the
  upper panel, but for merger mass ratios $\xi\geq 0.1$, for critical
  overdensities $\Deltac=\Deltavir$ (solid histogram) and
  $\Deltac=20000$ (dot-dashed histogram). The dashed histogram
  represents the results obtained by \citetalias{Nip17} for encounters
  at $\rcen\approx 0.02\rvir$, using simulations of satellites in
  isolated host haloes.}
\label{fig:ecc}
\end{figure}
%%%%%%%%%%%%%%%%%%%%%%%

As mentioned in Section~\ref{sec:sample}, we do not expect to have a
rapid merger for all halo-halo encounters.  Rapid mergers occur when
the orbits are bound ($\Etwob<0$), but also for unbound orbits
($\Etwob\geq 0$), provided the orbital angular-momentum modulus $L$ is
sufficiently low \citep[see section 7.4 of][]{Bin87}.  For this
reason, a convenient parameter that can be used to identify mergers is
the orbit eccentricity
\begin{equation}
e = \sqrt{1+\frac{2\Etwob L^2}{G^2\Mtwob^2}},
\end{equation}
which, for $\Etwob>0$, is an increasing function of both $\Etwob$ and
$L$. As in \citetalias{Nip17}, we take as fiducial discriminating
value of eccentricity $\ecrit=1.5$ and classify an encounter as a {\it
  merger} when $e \leq \ecrit$ and as a {\em fly-by} when
$e>\ecrit$. The eccentricity distributions for our samples of
halo-halo encounters with mass ratio $0.01\leq \xi<0.1$ are shown in
\Fig\ref{fig:ecc} (upper panel) for $\Deltac=\Deltavir$,
$\Deltac=5000$ and $\Deltac=10000$ (corresponding to radii $\rvir$,
$\rfivetc\approx0.15\rvir$ and $\rtentc\approx0.1\rvir$, respectively;
see \Sect\ref{sec:masses_radii}). The distribution of the eccentricity
for encounters with mass ratio $\xi\geq 0.1$ is shown in the lower
panel of \Fig\ref{fig:ecc} for $\Deltac=\Deltavir$ and $\Deltac=20000$
(corresponding to radii $\rvir$ and $\rtwentytc\approx0.06\rvir$,
respectively; see \Sect\ref{sec:masses_radii}). In the same panel we
plot also the distribution found by \citetalias{Nip17} for numerical
models of satellites orbiting in isolated haloes with $\xi\simeq0.13$
and $\xi\simeq 0.67$, measured at $\rcen\approx0.02\rvir$ (see
\Sect\ref{sec:masses_radii}).  From \Fig\ref{fig:ecc} it is clear that
most of the encounters are indeed classified as mergers: the adopted
cut in eccentricity allows us to effectively exclude the tail of
high-eccentricity orbits, which are most likely fly-bys.  The number
of mergers and the total number of encounters for our sample are
reported in \Tab\ref{tab:encounters} for different values of $\Deltac$
and intervals of $\xi$.

As a quantitative test of our classification of mergers and fly-bys,
we analysed the post-encounter evolution of the satellites in the
$\Deltac=20000$ catalogue with mass ratio $\xi\geq 0.1$. In practice,
for each encounter occurring between the snapshots at redshifts
$z_{i-1}$ and $z_{i}$, we check whether the satellite and the main
halo are distinct (i.e.\ the satellite has escaped) in the snapshot at
redshift $z_{i+1}$ (clearly we exclude the case $i=13$, because the
snapshot at $z_{13}=0$ is the last; see \Sect\ref{sec:sample}). We
find that the satellite escapes in 80\% of the encounters classified
as fly-bys and in 15\% of the encounters classified as mergers, which
suggests that our classification is sufficiently accurate.  We
verified that the selection of mergers is not sensitive to the exact
value of $\ecrit$: the main results of the present work are
essentially the same for values of $\ecrit$ in the range $1.25\lesssim
\ecrit \lesssim 2$.

\begin{table}
%\begin{center}
  \caption{Total number of encounters and number of encounters
    classified as mergers experienced by all the haloes in our sample
    (see \Sect\ref{sec:sample}). The data for $\rDelta=\rcen$ refer to
    the results of \citetalias{Nip17}.}
  \label{tab:encounters}
\begin{tabular}{clrr}
\hline
$\xi$ & $\rDelta$ &  Encounters & Mergers\\
\hline
$0.005-0.05$ &   $\rtwohc$ &   1855 & 1733 \\
$0.01-0.1$ &   $\rvir$ &   1049 & 998 \\
$0.01-0.1$ & $\rfivetc$ &  456 & 330 \\
$0.01-0.1$ & $\rtentc$ &  275 & 210 \\
$0.1-1$ &   $\rvir$ &   216 & 207 \\
$0.1-1$ &  $\rtwentytc$ &  98 & 69 \\
$0.1-1$ & $\rcen$ &  82 & 44 \\
\hline
\end{tabular}
%\end{center}
\end{table}

\subsection{Distribution of merger mass ratio}
\label{sec:mass_ratio}

%%%%%%%%%%%%%%FIG 2
\begin{figure}
  \centerline{\psfig{file=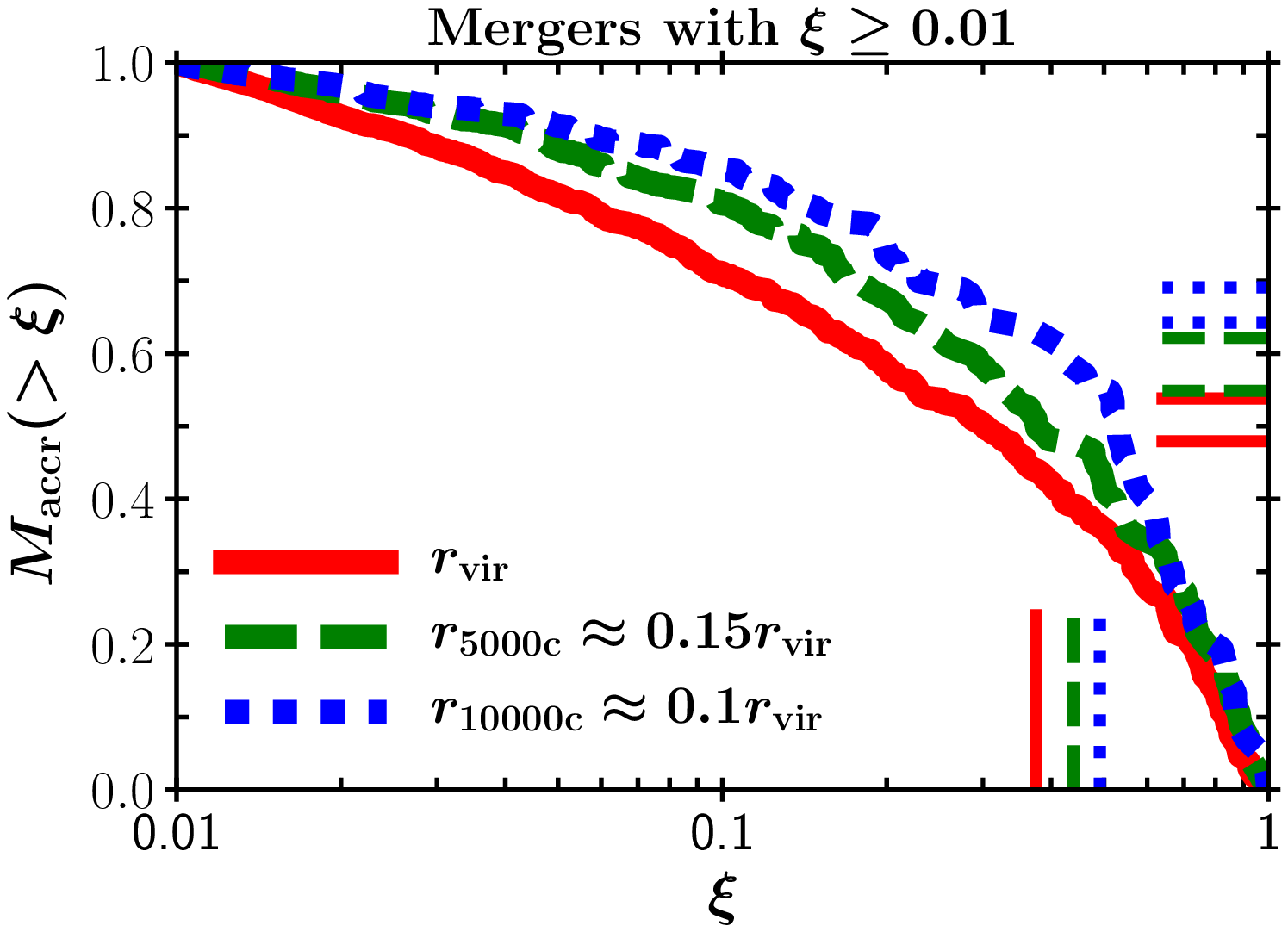,width=\hsize}}
  \centerline{\psfig{file=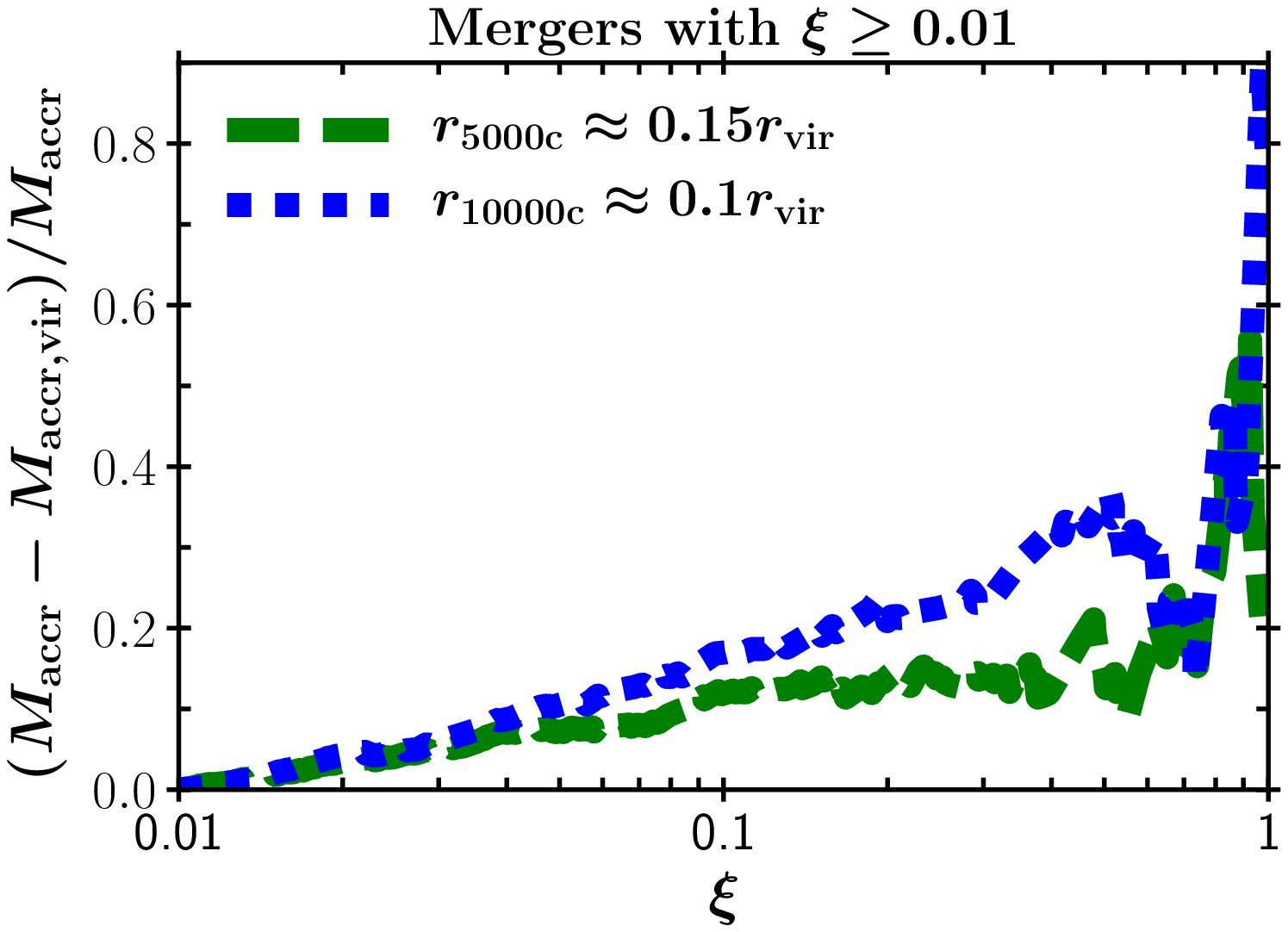,width=\hsize}}  
    \caption{{\em Upper panel.} Fraction of mass accreted in mergers
      with mass ratio larger than $\xi$, relative to the total mass
      accreted in mergers with $0.01\leq\xi\leq1$, for critical
      overdensities $\Deltac=\Deltavir$ (solid curve), $\Deltac=5000$
      (dotted curve) and $\Deltac=10000$ (dashed curve). The measures
      are for mergers in the redshift interval $0<z<1$ for our sample
      of dark-matter haloes with $\Mvir\geq10^{14}\Msun$ at $z=0$.
      The horizontal lines indicate, for the distributions with the
      corresponding line styles, the fraction $\fmajor$ of mass
      accreted in major mergers, assuming major-merger mass-ratio
      threshold $\ximajor=1/3$ (lower lines) or $\ximajor=1/4$ (upper
      lines).  The vertical lines indicate the mass-weighted average
      merger mass ratio (\Eq\ref{eq:mass-weighted}) for the
      distributions with the corresponding line styles. {\em Lower
        panel.}  Dotted curve: relative difference between the dotted
      and solid curves in the upper panel. Dashed curve: relative
      difference between the dashed and solid curve in the upper
      panel.}
\label{fig:mass_ratio}
\end{figure}
%%%%%%%%%%%%%%%%%%%%%%%

There are good reasons to expect mergers onto CGs in clusters to be
characterised by a distribution of mass ratios $\xi$ different from
that of cosmological halo-halo mergers measured at the virial radius.
It is well known that dynamical friction, which is the main driver of
galactic cannibalism, is more effective for more massive satellites,
so we expect the typical mass ratio of mergers onto central galaxies
to be higher than that of mergers at the virial radius of the host
cluster.  We can quantitatively explore this question by comparing the
distributions of $\xi$ in halo catalogues obtained for different
values of $\Deltac$.

The upper panel of \Fig\ref{fig:mass_ratio} shows, as a function of
the mass ratio $\xi$, the fraction $\Maccr(>\xi)$ of mass accreted in
mergers with mass ratio larger than $\xi$, normalised to the total
mass accreted in mergers with mass ratio $\xi\geq 0.01$, for mergers
measured at\footnote{Here we do not consider $\rtwotc$, because the
  corresponding sample of mergers has $\ximin>0.01$ (see
  \Sect\ref{sec:sample}).} $\rvir$, $\rfivetc$ and $\rtentc$. A clear
trend emerges from this plot: in line with the expectations, mergers
with higher mass ratios contribute more when more central halo regions
are considered. The difference between the mergers measured at
$\Deltavir$ and those measured at higher overdensities becomes more
and more important for $\xi \to 1$ (see lower panel of
\Fig\ref{fig:mass_ratio}).  The median value $\ximed$, such that half
of the mass is accreted in mergers with mass ratio larger than
$\ximed$, is 0.26 at $\rvir$, 0.37 at $\approx0.15\rvir$ and 0.53 at
$\approx0.1\rvir$.  Another useful indicator of the characteristic
mass ratio of mass accretion is the mass-weighted merger mass ratio
$\avxiM$ (see \citealt{Nip12}), which can be written as
\begin{equation}
\avxiM=\frac{\av{\xi^2}_N}{\avxiN},
\label{eq:mass-weighted}
\end{equation}
where $\av{\cdots}_{N}$ is the number-weighted average.  As shown in
\Fig\ref{fig:mass_ratio} (upper panel), $\avxiM\simeq 0.38$ at
$\rvir$, $\avxiM\simeq 0.44$ at $\approx 0.15\rvir$, and $\avxiM\simeq 0.49$ at $\approx 0.1\rvir$.  For the innermost radius here probed ($\approx 0.1\rvir$) the characteristic merger mass ratio is close to $1/2$.

Given a discriminant mass ratio $\ximajor$ between major and minor
mergers, we can define $\fmajor\equiv\Maccr(\geq \ximajor)/\Maccr(\geq
\ximin)$ as the fraction of mass accreted in major mergers in the
redshift range $0<z<1$ (here the minimum mass ratio is
$\ximin=0.01$). For, respectively, $\Deltac=\Deltavir$, $5000$ and
$10000$ we find $\fmajor=0.48$, $0.55$ and $0.64$ (assuming
$\ximajor=1/3$), and $\fmajor=0.54$, $0.62$ and $0.69$ (assuming
$\ximajor=1/4$).  Taking the results for $\Deltac=10000$ as a proxy
for accretion onto the CG, we can conclude that (at least in the
explored mass ratio interval $0.01\leq\xi\leq1$) {\em more than $60\%$
  of the mass accreted at $z<1$ by CGs in clusters is due to major
  mergers.} This conclusion is qualitatively consistent with previous
observational \citep{Lid13} and theoretical \citep{Rod16} results on
the role of major mergers in the build-up of massive CGs.

By definition, for given $\ximajor$, $\fmajor$ depends on the minimum
mass ratio $\ximin$.  Here, for the reasons explained in
\Sect\ref{sec:sample}, we have fixed $\ximin=0.01$, but of course also
mergers with lower mass ratio contribute to the actual halo mass
growth. The slopes at low $\xi$ of the curves in the upper panel of
\Fig\ref{fig:mass_ratio} suggest that the relative contribution of
mergers with $\xi<0.01$ is more important at $\rvir$ than at
$\rfivetc$ and $\rtentc$. To quantify this effect, we computed
$\fmajor$ assuming $\ximin=0.005$ (thus relaxing our requirement that
the satellites have at least $N\approx 100$ particles): in this case
we get values of $\fmajor$ that are only slightly smaller then those
obtained for $\ximin=0.01$ (for instance, by $\lesssim 6\%$ for
measures at $\rvir$ and by $\lesssim3\%$ for measures at
$\rtentc$). Thus, in this respect, our conclusion about the
predominance of major mergers in the $z\lesssim1$ build-up of cluster
CGs appears robust.

%  $\fmajor=0.45$ for
%  $\Deltac=\Deltavir$, $\fmajor=0.53$ for $\Deltac=5000$ and
%  $\fmajor=0.63$ for $\Deltac=10000$, when $\ximajor=1/3$, and
%  $\fmajor=0.51$ for $\Deltac=\Deltavir$, $\fmajor=0.60$ for
%  $\Deltac=5000$ and $\fmajor=0.67$ for $\Deltac=10000$, when
%  $\ximajor=1/4$. These

\subsection{Orbital parameters for mergers with mass ratio $\boldsymbol{0.01\leq\xi<0.1}$}
\label{sec:minor}

%%%%%%%%%%%%%%FIG 3
\begin{figure*}
  \centerline{\psfig{file=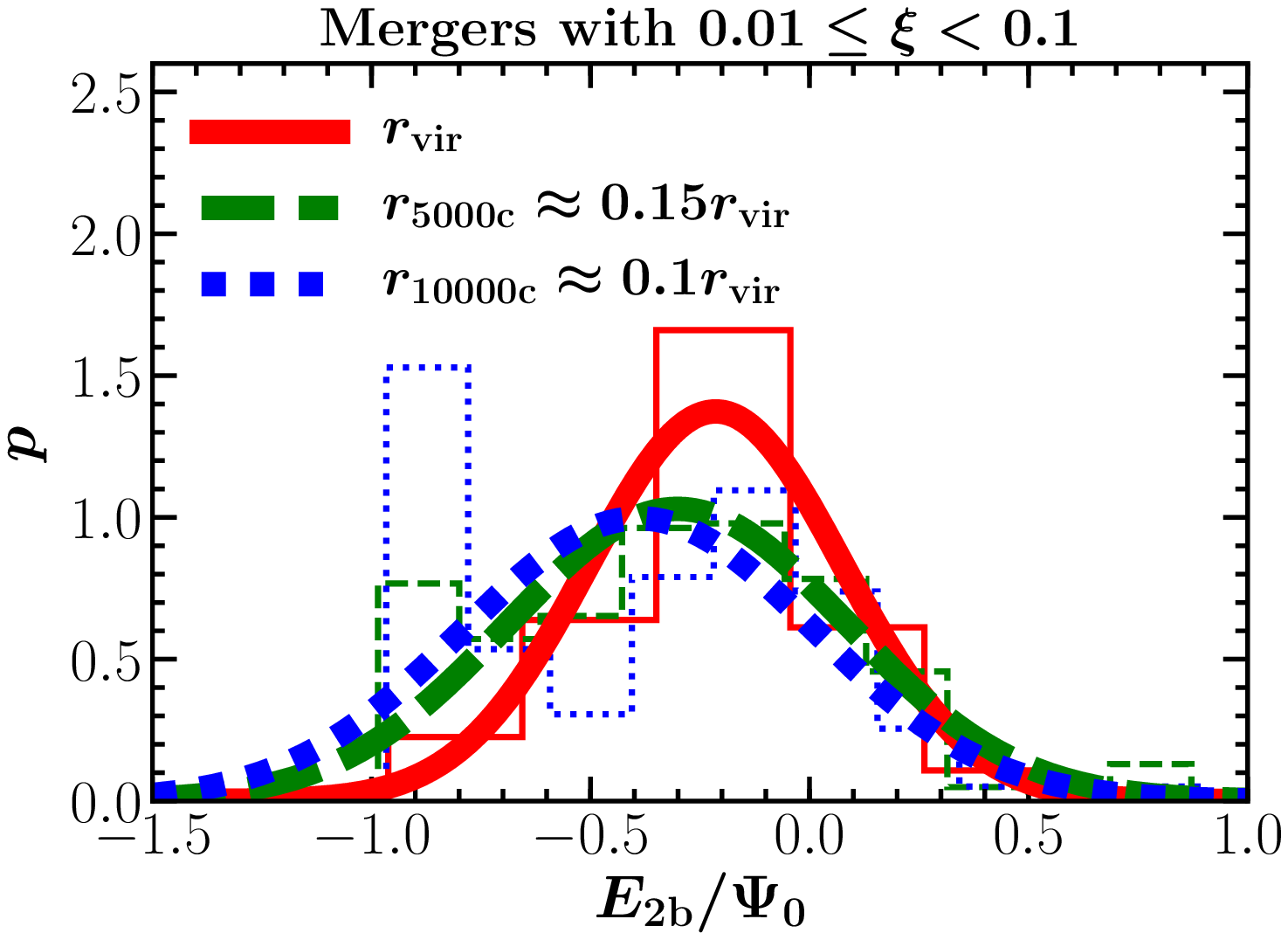,width=0.5\hsize}\psfig{file=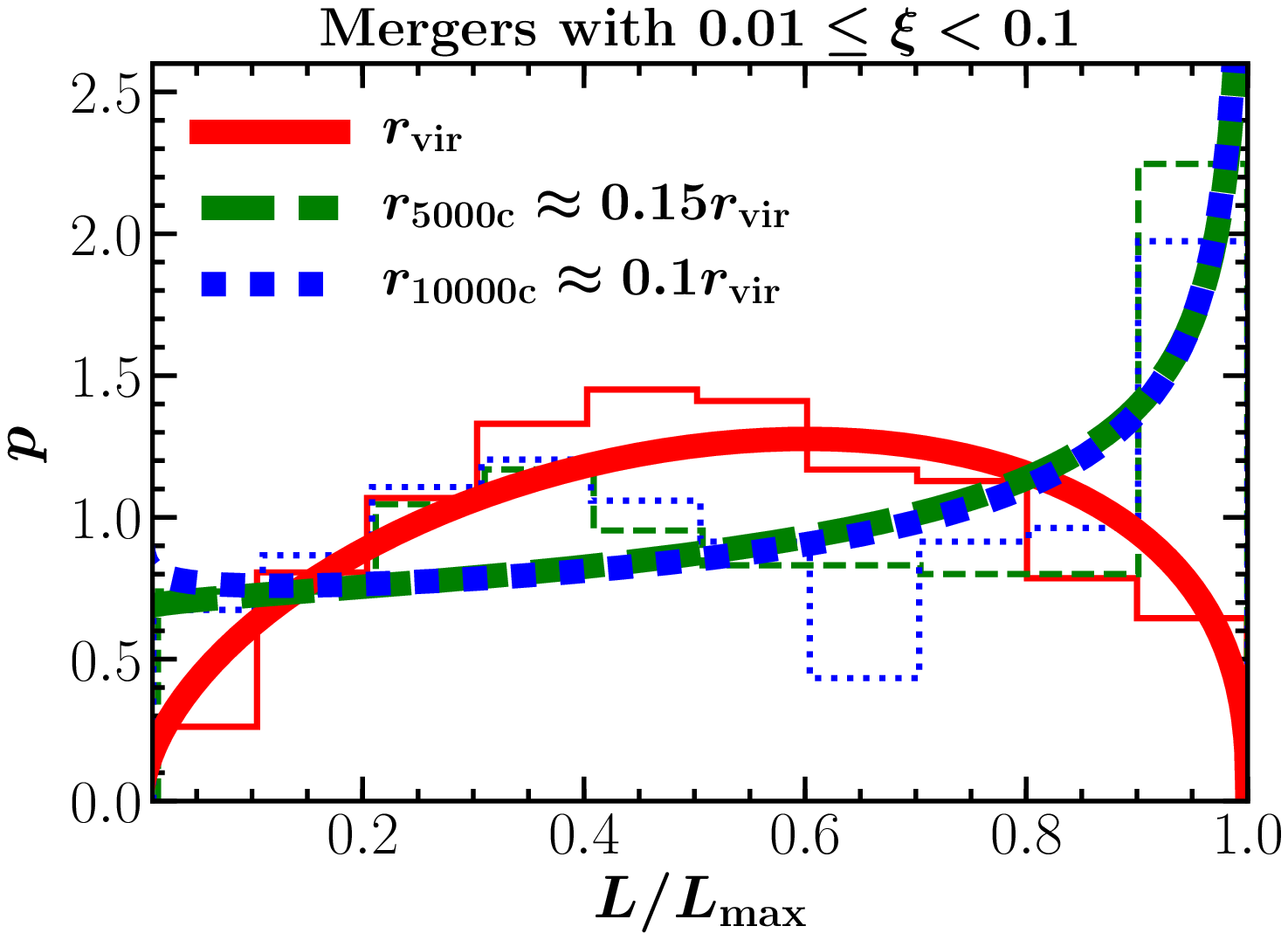,width=0.5\hsize}}
\caption{Probability distribution $p=\d n/\d x$ of the normalised
  orbital energy, computed in the two-body approximation
  ($x=\Etwob/\Psizero$; left-hand panel), and of the normalised
  angular-momentum modulus ($x=L/\Lmax$; right-hand panel) for
  critical overdensities $\Deltac=\Deltavir$ (solid histogram),
  $\Deltac=5000$ (dashed histogram) and $\Deltac=10000$ (dotted
  histogram).  Here $\Psizero\equiv G
  (\MDeltahost+\MDeltasat)/\rDeltahost$ and $\Lmax$ is defined by
  \Eq(\ref{eq:lmax}). The curves represent the best-fitting
  distributions of the histograms with the corresponding line styles
  (see \Tabs\ref{tab:gauss} and \ref{tab:beta}). Here we consider
  mergers with mass ratios $0.01\leq \xi< 0.1$.}
\label{fig:el_delta}
\end{figure*}
%%%%%%%%%%%%%%%%%%%%%%%

%%%%%%%%%%%%%%FIG 4
\begin{figure*}
\centerline{\psfig{file=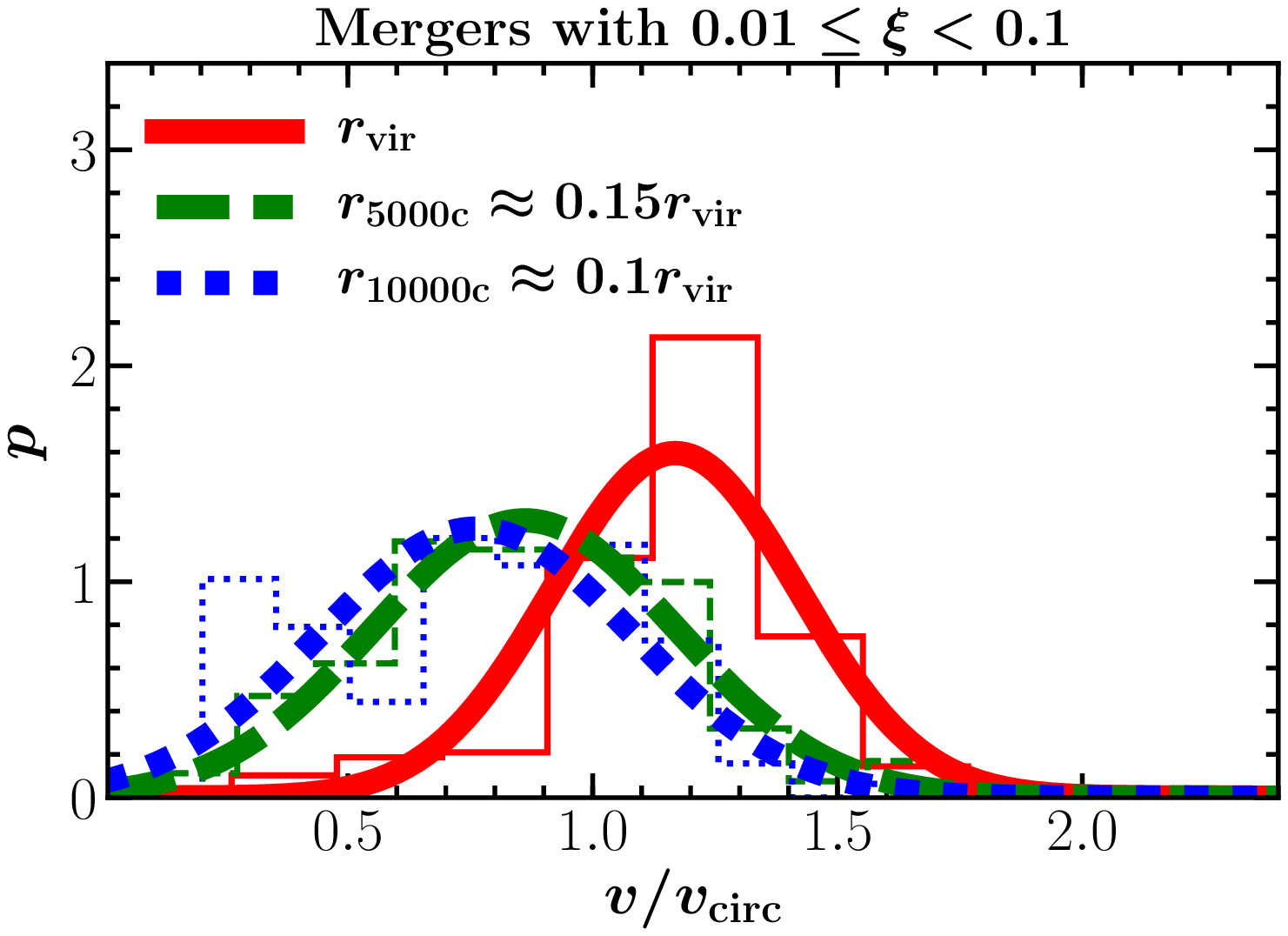,width=0.5\hsize}\psfig{file=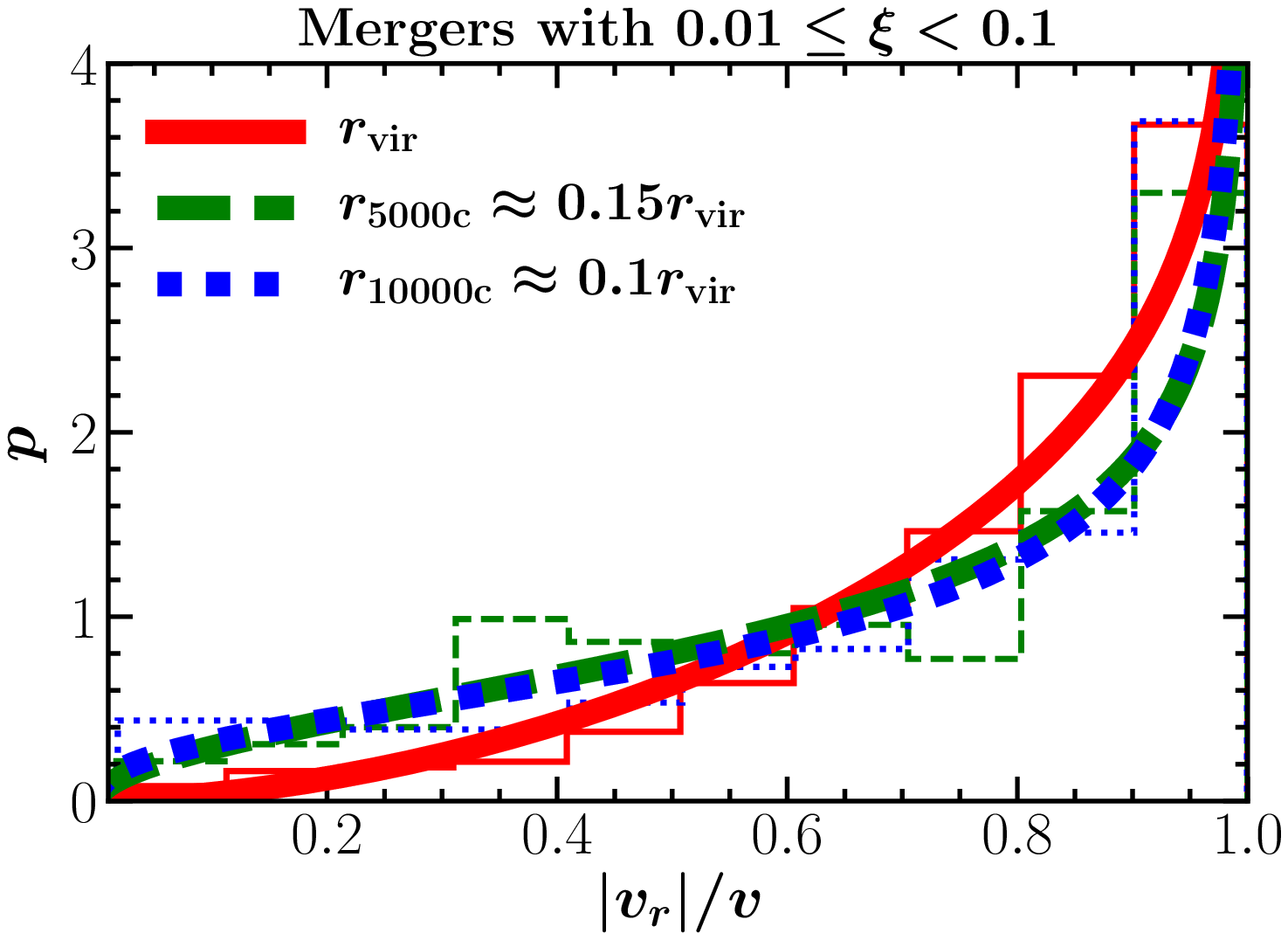,width=0.5\hsize}}
\caption{Probability distribution $p=\d n/\d x$ of the relative speed
  ($x=v/\vcirc$; left-hand panel) and of the radial-to-total relative
  velocity ratio ($x=|\vr|/v$; right-hand panel) when the satellite
  crosses the virial radius of the host $\rDeltahost$ for critical
  overdensities $\Deltac=\Deltavir$ (solid curve), $\Deltac=5000$
  (dashed curve) and $\Deltac=10000$ (dotted curve). Here $\vcirc$ is
  the host circular velocity at $\rDeltahost$.  $v/\vcirc$ and
  $|\vr|/v$ are evaluated at $\rDeltahost$ as in \Eqs(\ref{eq:vcorr})
  and (\ref{eq:vrcorr}).  The curves represent the best-fitting
  distributions of the histograms with the corresponding line styles
  (see \Tabs\ref{tab:gauss} and \ref{tab:beta}). Here we consider
  mergers with mass ratios $0.01\leq \xi< 0.1$.}
\label{fig:vvr_delta}
\end{figure*}
%%%%%%%%%%%%%%%%%%%%%%%

In this section we discuss the distribution of orbital parameters for
mergers (i.e.\ encounters with eccentricity $e\leq 1.5$) with mass
ratio in the range $0.01\leq\xi<0.1$, comparing the results for
$\Deltac=\Deltavir$, $\Deltac=5000$ and $\Deltac=10000$.
\Fig\ref{fig:el_delta} shows, for these samples of mergers, the
distributions of the two-body specific orbital energy $\Etwob$ and of
the modulus of the specific orbital angular momentum $L$. $\Etwob$ is
normalised to $\Psizero\equiv G (\MDeltahost+\MDeltasat)/\rDeltahost$,
which is the absolute value of the two-body gravitational potential of
the encounter when the separation is $\rDeltahost$. $L$ is normalised
to $\Lmax$ (\Eq\ref{eq:lmax}), which is the modulus of the maximum
angular momentum for given orbital energy $\Etwob$. We have fitted the
distributions of $\Etwob/\Psizero$ with a Gaussian distribution
\begin{equation}
p(x)=\frac{1}{\sqrt{2\pi}\sigma}\exp{\left[-\frac{(x-\mu)^2}{2\sigma^2}\right]},
\label{eq:gauss}
\end{equation}
where $\mu$ is the mean and  $\sigma$ is the standard deviation,
and the distributions of $L/\Lmax$ with a beta distribution
\begin{equation}
p(x)=\frac{x^{\alpha-1}(1-x)^{\beta-1}}{B(\alpha,\beta)},
\label{eq:beta}
\end{equation}
where
\begin{equation}
B(\alpha,\beta)=\frac{\Gamma(\alpha)\Gamma(\beta)}{\Gamma(\alpha+\beta)}
\end{equation}
and $\Gamma$ is the gamma function.  In \Fig\ref{fig:vvr_delta} we
plot the distributions of the orbital parameters $v/\vcirc$ and
$|\vr|/v$, which, as pointed out in \Sect\ref{sec:orbital_parameters},
are a pair of parameters, alternative to $\Etwob$ and $L$, often used
to characterise the orbits of galaxy and halo encounters. We emphasise
that, in the present context, this pair of parameters does not carry
exactly the same information as $E_{2b}$ and $L$: $v/\vcirc$ and
$|\vr|/v$ are evaluated at a separation $\rDeltahost$
(\Eqs\ref{eq:vcorr} and \ref{eq:vrcorr}), while $\Etwob$ and $L$ are
evaluated at the snapshot before the merger. Moreover, while $v$ is
normalised to the main halo circular velocity $\vcirc$, which is
independent of the properties of the satellite, $\Etwob$ is normalised
to $\Psizero$, which depends also on the mass of the satellite (and
therefore on the mass ratio $\xi$). The distributions of $v/\vcirc$
are fitted with a Gaussian (\Eq\ref{eq:gauss}), while the
distributions of $|\vr|/v$ are fitted with a beta distribution
(\Eq\ref{eq:beta}).  The best fitting distributions of
$\Etwob/\Psizero$, $L/\Lmax$, $v/\vcirc$ and $|\vr|/v$ are
over-plotted in the corresponding panels of \Fig\ref{fig:el_delta} and
\Fig\ref{fig:vvr_delta}, and their parameters are reported in
\Tabs\ref{tab:gauss} and \ref{tab:beta}.

The distributions of $\Etwob/\Psizero$ (left-hand panel in
\Fig\ref{fig:el_delta}) suggest that for higher values of $\Deltac$
(i.e.\ when more central regions of the haloes are considered) the
orbits of mergers tend to be slightly more bound: the mean orbital
energy for $\Deltac=10000$ is more negative than that for
$\Deltac=5000$, which in turn is more negative than that at
$\Deltac=\Deltavir$ (see \Tab\ref{tab:gauss}). However, the
distributions are relatively broad and there is substantial overlap.
A qualitatively similar, but stronger trend can be seen in the
distributions of $v/\vcirc$ (which is another measure of the binding
energy of the orbit; left-hand panel in \Fig\ref{fig:vvr_delta}),
which are characterised by larger offsets between the peak of the
measures at $\rvir$ and those at $\rfivetc$ or $\rtentc$. The fact
that the distributions are more offset in $v/\vcirc$ than in
$\Etwob/\Psizero$ comes from the fact that the accretion history for
higher $\Deltac$ is characterised by higher merger mass ratios (see
\Sect\ref{sec:mass_ratio} and \Fig\ref{fig:mass_ratio}). As pointed
out above, while $\vcirc$ ignores the properties of the satellite, the
normalisation potential $\Psizero$ accounts for the mass ratio. In
this sense, $\Etwob/\Psizero$ should give a cleaner measure of the
binding energy of the orbits, when samples with different mass-ratio
distributions are compared. We also note that the distributions of
both $\Etwob/\Psizero$ and $v/\vcirc$ have higher scatter (larger
standard deviation; see \Tab\ref{tab:gauss}) for increasing $\Deltac$.

The distributions of $L/\Lmax$ (right-hand panel in
\Fig\ref{fig:el_delta}) and $|\vr|/v$ (right-hand panel in
\Fig\ref{fig:vvr_delta}) suggest that for higher values of $\Deltac$
(i.e.\ when more central regions of the haloes are considered) the
orbits of mergers tend to be significantly more tangential.  Comparing
the right-hand panels of \Figs\ref{fig:el_delta} and
\ref{fig:vvr_delta}, it is apparent that, for $\Deltac=5000$ and
$\Deltac=10000$, the distributions of both $L/\Lmax=\vtan/v$ and
$|\vr|/v$ peak at $\approx 1$. This might be counterintuitive, because
$|\vr|/v\to 0$ when $|\vtan|/v\to 1$. However, the relation between
$|\vr|/v$ and $|\vtan|/v$, namely
\begin{equation}
 \frac{|\vr|}{v}=\sqrt{1-\left(\frac{\vtan}{v}\right)^2},
\end{equation}
is non-linear and such that, for instance, $|\vr|/v\gtrsim 0.9$
corresponds to $\vtan/v\lesssim 0.44$. Thus, a peak at $|\vr|/v
\approx 1$ does not necessarily imply a peak at $L/\Lmax\approx
0$. Moreover, we recall that while $L/\Lmax$ is evaluated at $\rsnap$,
$|\vr|/v$ is evaluated at $\rDeltahost$, so the right-hand panels of
\Figs\ref{fig:el_delta} and \ref{fig:vvr_delta} do not contain exactly
the same information.

The distributions of $L/\Lmax$ and $|\vr|/v$ for $\Deltac=5000$ and
$\Deltac=10000$ are almost indistinguishable, but they are
significantly different from those obtained for $\Deltac=\Deltavir$.
For the higher overdensities, the mean values of $L/\Lmax$ and
$|\vr|/v$ are, respectively, higher and lower than those obtained for
$\Deltac=\Deltavir$ (see \Tab\ref{tab:beta}).  The difference between
measures at $\rvir$ and those at $\lesssim 0.15\rvir$ is best
visualised in the right-hand panel of \Fig\ref{fig:el_delta}: the
distribution of $L/\Lmax$ for measures at $\rvir$ drops for
$L/\Lmax\to 1$, where instead the distributions for more central
measures peak.  The values of the standard deviation reported in
\Tab\ref{tab:beta} indicate that the distributions of $L/\Lmax$ and
$|\vr|/v$ have larger scatter for $\Deltac=5000$ and $\Deltac=10000$
than for $\Deltac=\Deltavir$.  Overall, the results presented in this
section lead us to conclude that, for mergers with $0.01\leq\xi<0.1$,
{\em the orbits of mergers at $\rtentc\approx 0.1\rvir$ are more bound
  and more tangential than those at $\rvir$.}

\begin{table}
%\begin{center}
  \caption{Mean ($\mu$) and standard deviation ($\sigma$) of the
    best-fitting Gaussian distributions (\Eq\ref{eq:gauss}) of the
    orbital parameters $\Etwob/\Psizero$ and $v/\vcirc$, for different
    values of the overdensity radius $\rDelta$ and intervals of mass
    ratios $\xi$. The data for $\rDelta=\rcen$ refer to the results of
    \citetalias{Nip17}.}
  \label{tab:gauss}
\begin{tabular}{lclrr}
\hline
Parameter &$\xi$ & $\rDelta$ &  $\mu$ & $\sigma$\\
\hline
$\Etwob/\Psizero$ &$0.01-0.1$ &   $\rvir$ &   -0.21 & 0.29 \\
$\Etwob/\Psizero$ & $0.01-0.1$ & $\rfivetc$ &  -0.30 & 0.39 \\
$\Etwob/\Psizero$  & $0.01-0.1$ & $\rtentc$ &  -0.39 & 0.40 \\
$\Etwob/\Psizero$ &$0.1-1$ &   $\rvir$ &   -0.24 & 0.23 \\
$\Etwob/\Psizero$ & $0.1-1$ &  $\rtwentytc$ &  -0.31 & 0.37 \\
$\Etwob/\Psizero$  & $0.1-1$ & $\rcen$ &  -0.33 & 0.31 \\
$v/\vcirc$ & $0.01-0.1$  & $\rvir$ &  1.17 & 0.25 \\
$v/\vcirc$ & $0.01-0.1$  & $\rfivetc$  &  0.86 & 0.31 \\
$v/\vcirc$ & $0.01-0.1$  & $\rtentc$  &  0.76 & 0.32 \\
$v/\vcirc$ & $0.1-1$  & $\rvir$ &  1.28 & 0.21 \\
$v/\vcirc$ & $0.1-1$  & $\rtwentytc$  &  0.99 & 0.39 \\
$v/\vcirc$ & $0.1-1$  & $\rcen$  &  1.39 & 0.35 \\
\hline
\end{tabular}
%\end{center}
\end{table}

\begin{table}
%\begin{center}
 \caption{Parameters $\alpha$ and $\beta$, mean ($\mu$) and standard
   deviation ($\sigma$) of the best-fitting beta distributions
   (\Eq\ref{eq:beta}) of the orbital parameters $L/\Lmax$ and
   $\vr/v$, for different values of the overdensity radius $\rDelta$
   and intervals of mass ratios $\xi$. The data for $\rDelta=\rcen$
   refer to the results of \citetalias{Nip17}.}
 \label{tab:beta}
\begin{tabular}{lclrrrr}
\hline
Parameter &$\xi$ & $\rDelta$ & $\alpha$ & $\beta$ &  $\mu$ & $\sigma$\\
\hline
$L/\Lmax$ &$0.01-0.1$ &   $\rvir$ & 1.58 & 1.38 &  0.53 & 0.25 \\
$L/\Lmax$ & $0.01-0.1$ & $\rfivetc$ & 1.01 & 0.71 & 0.59 & 0.30 \\
$L/\Lmax$  & $0.01-0.1$ & $\rtentc$ & 0.96 & 0.69 & 0.58 & 0.30\\
$L/\Lmax$ &$0.1-1$ &   $\rvir$ & 1.57 & 2.16 & 0.42 & 0.23 \\
$L/\Lmax$ & $0.1-1$ &  $\rtwentytc$ & 0.81 & 0.85 & 0.49 & 0.31 \\
$L/\Lmax$  & $0.1-1$ & $\rcen$ & 1.99 & 0.68 & 0.75 & 0.23 \\
$\vr/v$ & $0.01-0.1$  & $\rvir$ & 2.69 & 0.78 & 0.78 & 0.20 \\
$\vr/v$ & $0.01-0.1$  & $\rfivetc$  & 1.51 & 0.67&  0.69 & 0.26 \\
$\vr/v$ & $0.01-0.1$  & $\rtentc$  &  1.39 & 0.60 & 0.70 & 0.26 \\
$\vr/v$ & $0.1-1$  & $\rvir$ &  3.70 & 0.67 & 0.85 & 0.16 \\
$\vr/v$ & $0.1-1$  & $\rtwentytc$  & 1.24 & 0.43 &  0.75 & 0.27 \\
$\vr/v$ & $0.1-1$  & $\rcen$  & 1.28 & 1.05 & 0.55 & 0.27 \\
\hline
\end{tabular}
%\end{center}
\end{table}

\subsection{Orbital parameters for mergers with mass ratio $\boldsymbol{\xi\geq 0.1}$}
\label{sec:major}

%%%%%%%%%%%%%%FIG 5
\begin{figure*}
  \centerline{\psfig{file=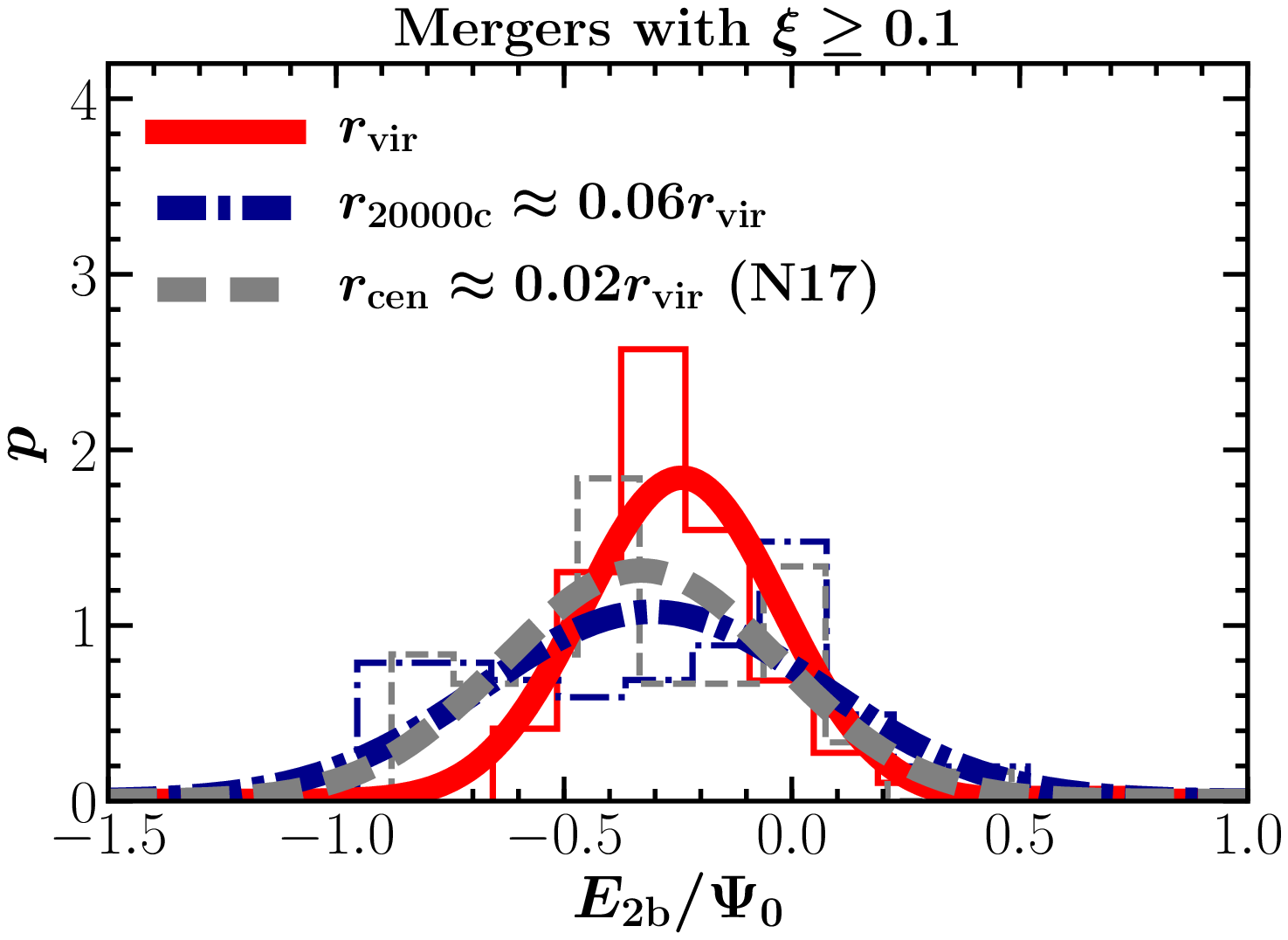,width=0.5\hsize}\psfig{file=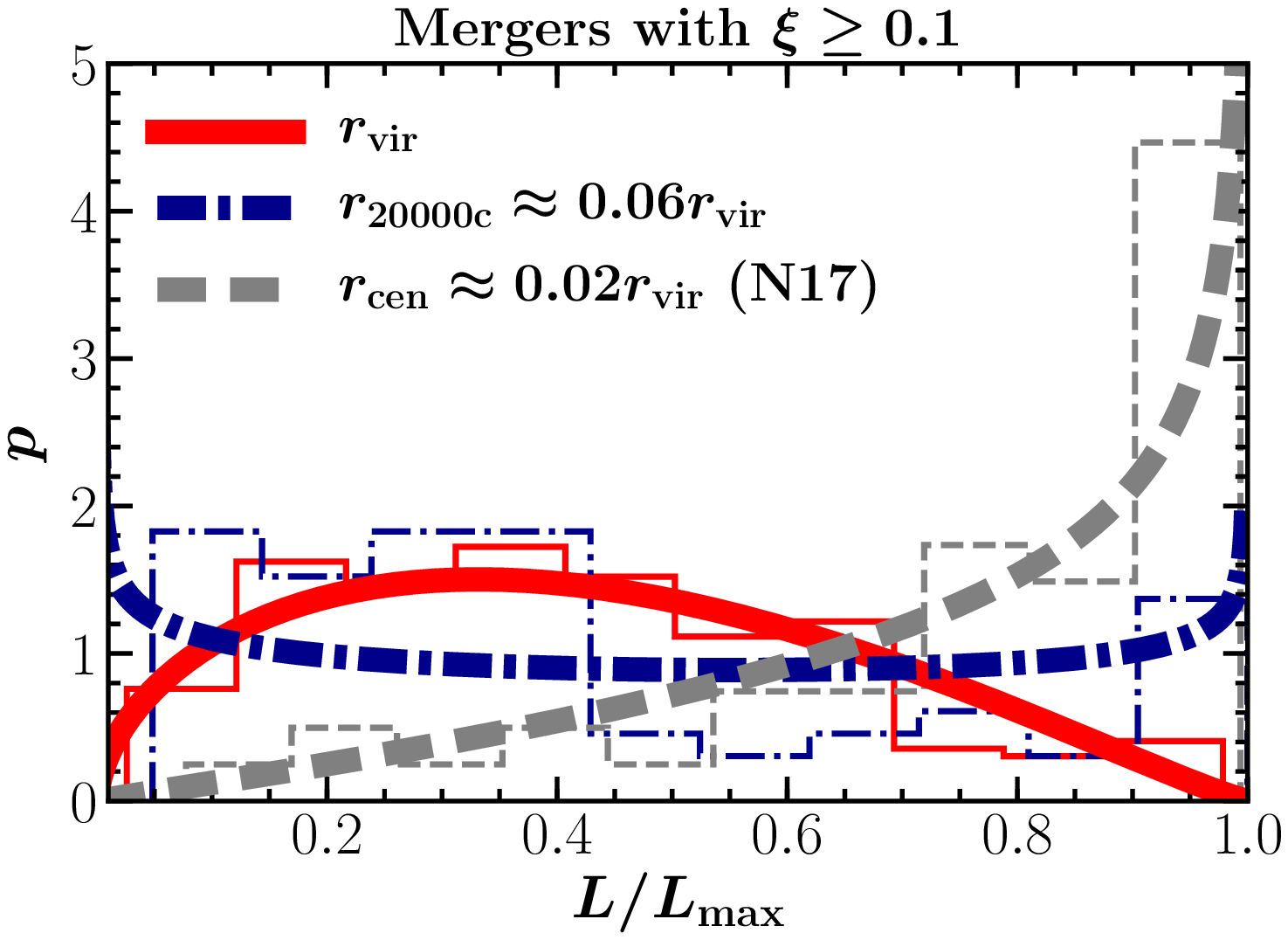,width=0.5\hsize}}
  \caption{Same as \Fig\ref{fig:el_delta}, but for merger mass ratios
    $\xi\geq 0.1$, for critical overdensities $\Deltac=\Deltavir$
    (solid curve) and $\Deltac=20000$ (short-dashed curve). The
    long-dashed curve represents the results obtained by
    \citetalias{Nip17} for mergers at $\rcen\approx 0.02\rvir$, using
    simulations of satellites orbiting in isolated host haloes.}
\label{fig:el_cen}
\end{figure*}
%%%%%%%%%%%%%%%%%%%%%%%

%%%%%%%%%%%%%%FIG 6
\begin{figure*}
  \centerline{\psfig{file=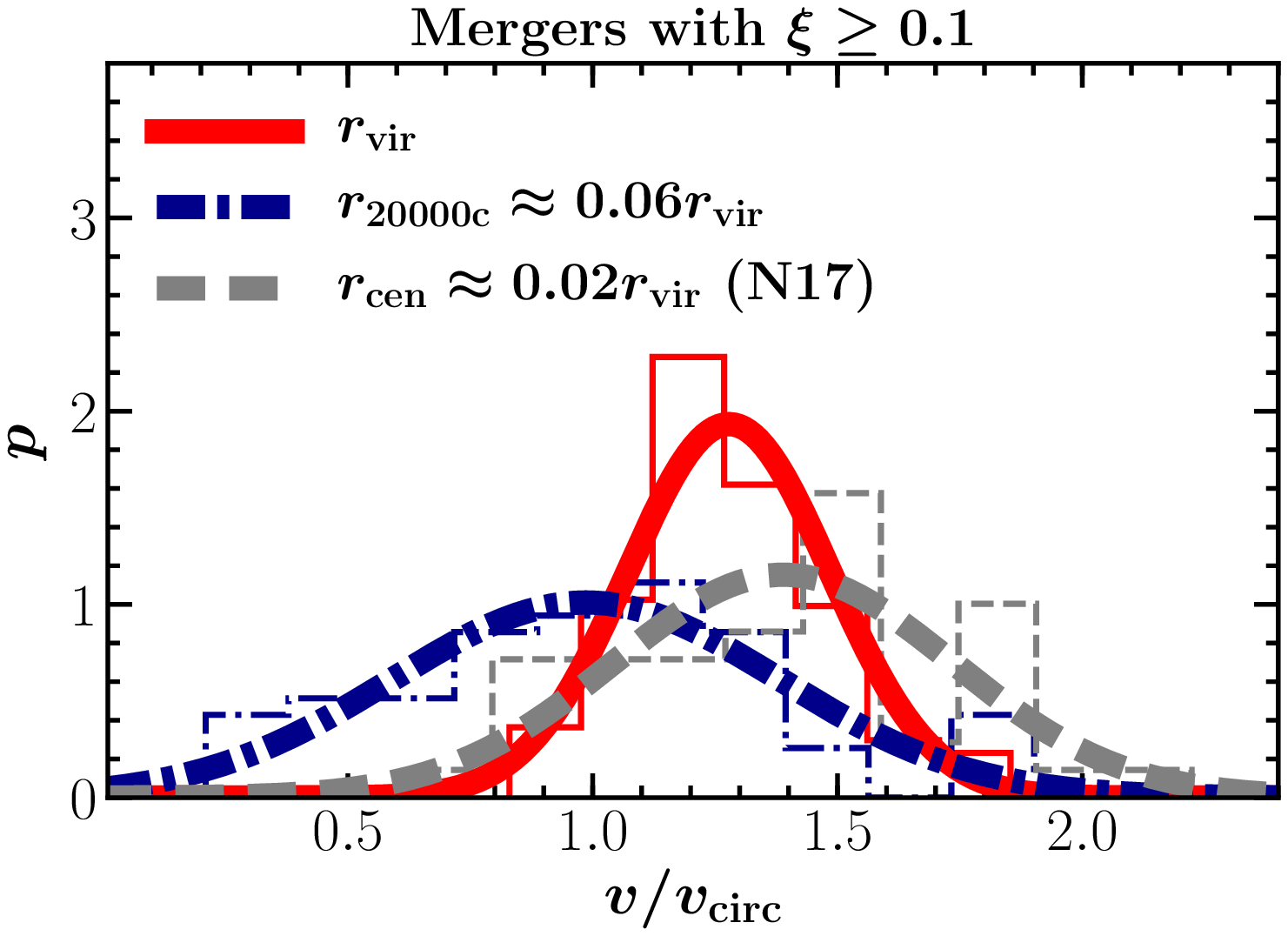,width=0.5\hsize}\psfig{file=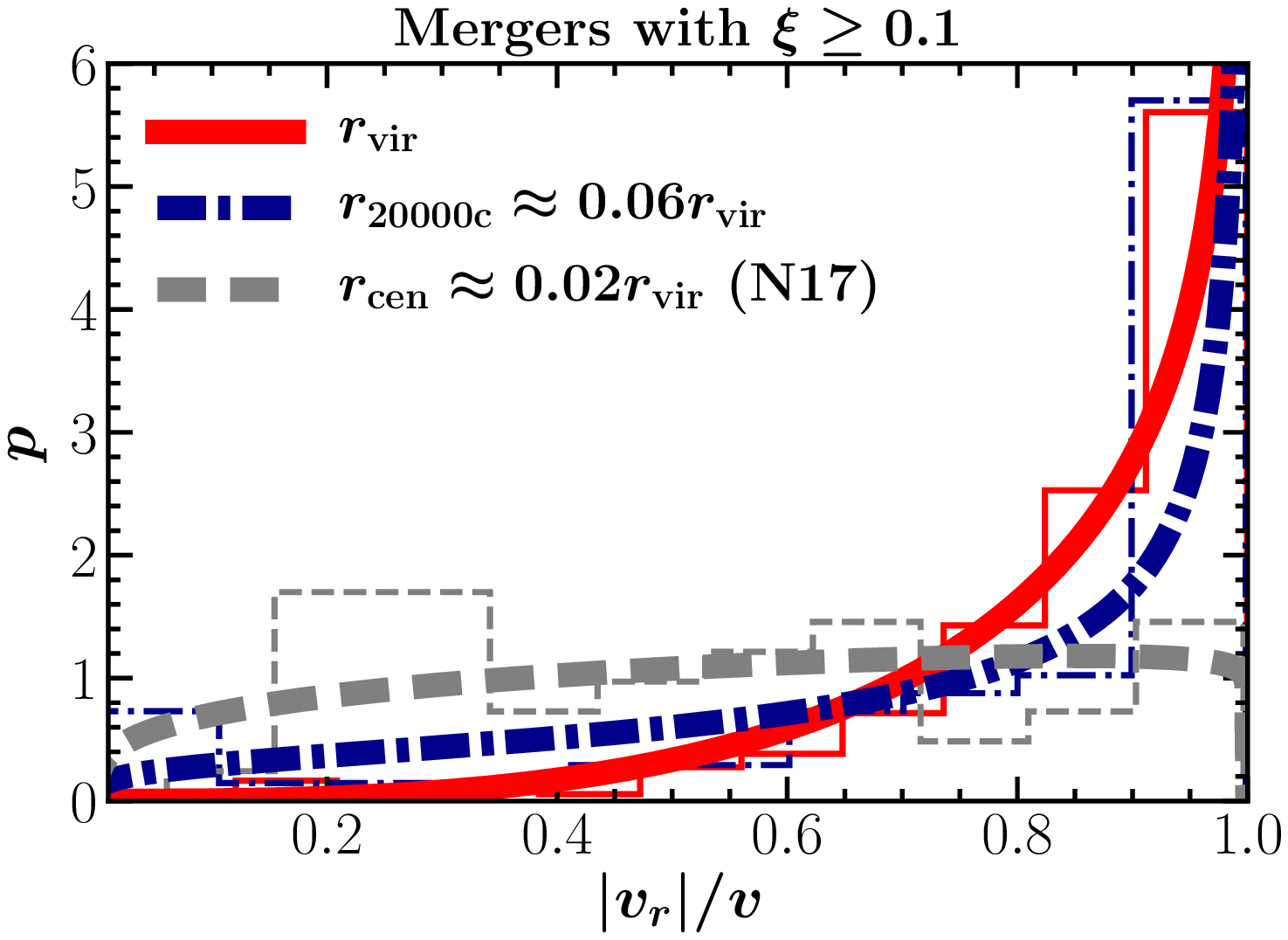,width=0.5\hsize}}
    \caption{Same as \Fig\ref{fig:vvr_delta}, but for merger mass
      ratios $\xi \geq 0.1$, for critical overdensities
      $\Deltac=\Deltavir$ (solid curves), and $\Deltac=20000$
      (short-dashed curves). The long-dashed curves represent the
      results obtained by \citetalias{Nip17} for mergers at
      $\rcen\approx 0.02\rvir$, using simulations of satellites
      orbiting in isolated host haloes.}
\label{fig:vvr_cen}
\end{figure*}
%%%%%%%%%%%%%%%%%%%%%%%

We focus here on the distribution of orbital parameters for mergers
with mass ratios in the range $0.1\leq \xi \leq 1$.  As we limit
ourselves to the mass ratios higher than $\xi=0.1$, we can consider
here the overdensity $\Deltac=20000$ (see \Sect\ref{sec:sample}),
which probes the central region of the halo
$\rtwentytc\approx0.06\rvir$ (see \Sect\ref{sec:masses_radii}).  As
done in \Sect\ref{sec:minor}, we compare the results obtained for this
central halo region with those obtained at $\rvir$. Moreover, in this
case we can include in our analysis also the results of
\citetalias{Nip17}, who, using non-cosmological simulations, explored
the distribution of orbital parameters at $\rcen\approx 0.02\rvir$,
roughly corresponding to $\Deltac=50000$, for merger mass ratios
$\xi\simeq 0.13$ and $\xi\simeq 0.67$ (see
\Sects\ref{sec:masses_radii} and \ref{sec:mergers_flybys}). Both
$\rtwentytc\approx0.06\rvir$ and $\rcen\approx0.02\rvir$ can be
considered proxies for the characteristic size of the CG in a cluster
of galaxy (see \Sect\ref{sec:masses_radii}).

In \Figs\ref{fig:el_cen} and \ref{fig:vvr_cen} we plot the
distributions of $\Etwob/\Psizero$, $L/\Lmax$, $v/\vcirc$, and
$|\vr|/v$, together with their best fits. The parameters of the best
fits (Gaussian distributions---\Eq\ref{eq:gauss}---for
$\Etwob/\Psizero$ and $v/\vcirc$, and beta
distributions---\Eq\ref{eq:beta}---for $L/\Lmax$ and $|\vr|/v$) are
reported in \Tabs\ref{tab:gauss} and \ref{tab:beta}.  The distribution
of $\Etwob/\Psizero$ that we find at $\rtwentytc\approx0.06\rvir$ has
slightly more negative mean than the distribution measured at $\rvir$
(see right-hand-panel of \Fig\ref{fig:el_cen}), confirming the trend
found in \Sect\ref{sec:minor} for mergers with $\xi<0.1$: the orbits
of satellites accreting onto CGs tend to be slightly more bound than
those of satellites accreting at the virial radius of the host
cluster.  The same result is visualised in \Fig\ref{fig:vvr_cen}
(left-hand panel), showing that the distribution of $v/\vcirc$ at
$\rtwentytc$ peaks at lower values than the distribution of $v/\vcirc$
measured at $\rvir$.

It is interesting to compare the results at
$\rtwentytc\approx0.06\rvir$ with those obtained at
$\rcen\approx0.02\rvir$ by \citetalias{Nip17}. While the distributions
of $\Etwob/\Psizero$ are very similar for $\rcen$ and $\rtwentytc$,
the distributions of $v/\vcirc$ are significantly offset: in the
experiments of \citetalias{Nip17} the values of $v/\vcirc$ measured at
$\rcen$ tend to be higher than those measured at $\rvir$, in contrast
with the results obtained here for $\rtwentytc$. The different
behaviour between the distributions of $v/\vcirc$ and
$\Etwob/\Psizero$ can be explained as follows. Though both quantities
measure the binding energy of the orbit, as pointed out in
\Sect\ref{sec:minor} they are normalised quite differently: while
$\Psizero$ accounts for the merger mass ratio $\xi$, $\vcirc$ is
independent of $\xi$. The experiments of \citetalias{Nip17} have
average merger mass ratio $\av{\xi}_N\simeq0.62$ higher than our
samples of mergers at $\rvir$ and $\rtwentytc$ ($\av{\xi}_N\approx0.5$
in the interval $0.1\leq \xi \leq 1$). Moreover, we recall that in
this work $\Etwob$ is measured at the snapshot before merger, while
$v$ is corrected to be evaluated at $\rDeltahost$ (see
\Sect\ref{sec:orbital_parameters}): this is another source of
difference between the distributions of $\Etwob/\Psizero$ and
$v/\vcirc$. Based only on measures of $v/\vcirc$, compared to those of
\citetalias{Jia15} (measured at $\rtwohc$), \citetalias{Nip17}
concluded that the orbits for CG-satellite mergers tend to be less
bound than those of cosmological halo-halo mergers. The present
analysis reveals that the higher values of $v/\vcirc$ found in the
simulations of \citetalias{Nip17} at least partly reflect a bias in
the merger mass ratios, which tend to be higher than cosmologically
motivated values. In any case, the simulations here considered should,
in general, be more realistic than the idealised simulations of
\citetalias{Nip17}, thus we believe that the distribution of
$v/\vcirc$ here obtained for $\rtwentytc$ should be more
representative for real CGs than the distribution found for $\rcen$ in
\citetalias{Nip17}. Therefore, based on the results found for
$\Etwob/\Psizero$ and $v/\vcirc$, we can conclude that {\em the orbits
  of satellites accreting onto CGs in clusters tend to be more bound
  than the orbits of satellites accreting onto the host cluster-size
  haloes}.

The distributions of $L/\Lmax$ (right-hand panel in
\Fig\ref{fig:el_cen}) and $|\vr|/v$ (right-hand panel in
\Fig\ref{fig:vvr_cen}) indicate that, as it happens for mergers with
$\xi<0.1$, also for mass ratios $\xi\geq 0.1$ the orbits measured at
$\rtwentytc$ tend to be more tangential than those measured at $\rvir$
(see also \Tab\ref{tab:beta}). In particular, the probability density
function of $L/\Lmax$ is flatter for $\Deltac=20000$ than for
$\Deltac=\Deltavir$ (the latter peaks at $\approx 0.3$ and drops above
$\approx 0.8$); the probability density function of $|\vr|/v$ measured
at $\rtwentytc$ has a strong peak at $\approx 1$, which is absent for
measures at $\rvir$.  As far as the eccentricity of the orbits is
concerned, the results obtained by \citetalias{Nip17} are consistent
with those obtained here: the distributions of $L/\Lmax$ and $|\vr|/v$
measured at $\rcen$ are biased towards, respectively, high and low
values, even more than those found here for $\Deltac=20000$. Therefore
the present results confirm and strengthen the finding of
\citetalias{Nip17} that {\em the orbits of satellites accreting onto
  CGs tend to be more tangential than those of cosmological halo-halo
  accretion at the virial radius}.

As far as the scatter in the distributions is concerned, the trend for
$\xi\geq 0.1$ is the same as that for $\xi<0.1$: for all the
considered parameters ($\Etwob/\Psizero$, $v/\vcirc$, $L/\Lmax$,
$|\vr|/v$) the scatter is larger for higher values of $\Deltac$
(i.e.\ smaller radii; see \Tabs\ref{tab:gauss} and
\ref{tab:beta}). The effect is strongest for $v/\vcirc$, for which the
best-fitting probability density function has standard deviation
almost a factor of $2$ higher for measures at $\approx 0.06\rvir$ than
for measures at $\rvir$.  This is qualitatively in agreement with the
findings of \citetalias{Nip17}: the standard deviations for measures
at $\rcen$ are higher than those for measures at $\rvir$, with the
only exception of $L/\Lmax$, for which the scatter is the same in the
two cases (\Tabs\ref{tab:gauss} and \ref{tab:beta}).

\section{Conclusions}
\label{sec:conclusions}

In this paper we have used the results of the dark-matter only
cosmological $N$-body simulations Le SBARBINE \citep{Des16} to study
the statistical properties of mergers between central and satellite
galaxies in galaxy clusters. In particular we selected a sample of 101
cluster-size haloes at $z=0$ from the simulations Ada and Bice and
traced their merging history in the redshift interval $0<z<1$. We
constructed merger trees for different overdensities $\Deltac$. When
we use the virial overdensity [$\Deltac=\Deltavir$, with
  $100\lesssim\Deltavir(z)\lesssim 150$] we probe the accretion of
satellites at the cluster virial radius $\rvir$
\citep{giocoli08,giocoli10c}. When we use higher overdensities
($\Deltac=5000$, $10000$ and $20000$) we probe the accretion of
satellites in the central region of the cluster (at radii
$\rfivetc\approx0.15\rvir$, $\rtentc\approx0.1\rvir$ and
$\rtwentytc\approx0.06\rvir$ ), which can be considered as a proxy for
the accretion of satellite galaxies onto CGs.  We measured the
distributions of merger mass ratios and orbital parameters for these
merger histories.  The main results of this work are the following.
\begin{itemize}
\item Though minor mergers largely outnumber major mergers, the latter
  contribute to the mass accreted at $z<1$ at least as much as minor
  mergers, for all values of $\Deltac$. The mass-weighted merger mass
  ratio $\av{\xi}_M$ increases for increasing $\Deltac$, so major
  mergers are even more important for CGs than for the accretion at
  the cluster virial radius. In the mass-ratio interval
  $0.01\leq\xi\leq 1$, more than $60\%$ of the mass accreted by CGs at
  $z<1$ is due to major mergers ($\xi\geq 1/3$).
\item For higher overdensities (i.e.\ more central regions), the
  orbits of the accreting satellites tend to be less bound and more
  tangential. Therefore, the orbits of satellites accreting onto CGs
  are characterised by higher specific orbital angular momentum and
  lower specific orbital energy than orbits of halo accretion at the
  virial radius.
\item The scatter in the orbital parameters tends to be larger for
  accretion onto CGs than for accretion at the halo virial radius.  In
  this respect, the strongest effect is found for the distribution of
  $v/\vcirc$, which, for $\xi\geq 0.1$, has standard deviation almost
  a factor of 2 higher at $\approx 0.06\rvir$ than at $\rvir$.
\item We compared the results obtained at $\rtwentytc\approx0.06\rvir$
  in our cosmological simulations with those obtained by
  \citetalias{Nip17} at $\rcen\approx0.02$ in idealised
  non-cosmological simulations. We found good agreement on the
  distribution of orbital angular momentum, but we revised
  \citetalias{Nip17}'s conclusions on the binding energy of the
  orbits, which were somewhat biased by the non-cosmological setting. 
  \item We provided parameters of the analytic best-fitting
  distributions of the pairs of orbital parameters ($\Etwob$,$L$) and
  ($v/\vcirc$,$|\vr|/v$) for different values of $\Deltac$
  (\Tabs\ref{tab:gauss} and \ref{tab:beta}). The distributions
  obtained for $\Deltac=20000$ (i.e.\ measured at
  $\rtwentytc\approx0.06\rvir$) can be taken as reference for modeling
  accretion onto CGs in clusters. In particular, the provided analytic
  distributions could be included in models attempting to predict the
  evolution of the scaling relations of cluster CGs without resorting
  to hydrodynamic cosmological simulations
  \citep[e.g.][]{Ber11,Vol13,Sha15}.
\end{itemize}

\section*{Acknowledgements}

We would like to thank Giuseppe Tormen (and the Physics and Astronomy Department of Padova) who provided the computational resources to run the simulations.  CG acknowledges support from the Italian Ministry for Education, University and Research (MIUR) through the SIR individual grant SIMCODE, project number RBSI14P4IH.

\appendix

\section{Distributions of the orbital parameters measured at $\boldsymbol{\rtwohc}$}
\label{sec:orb_par_at_r200}

\begin{figure*}
   \centerline{\psfig{file=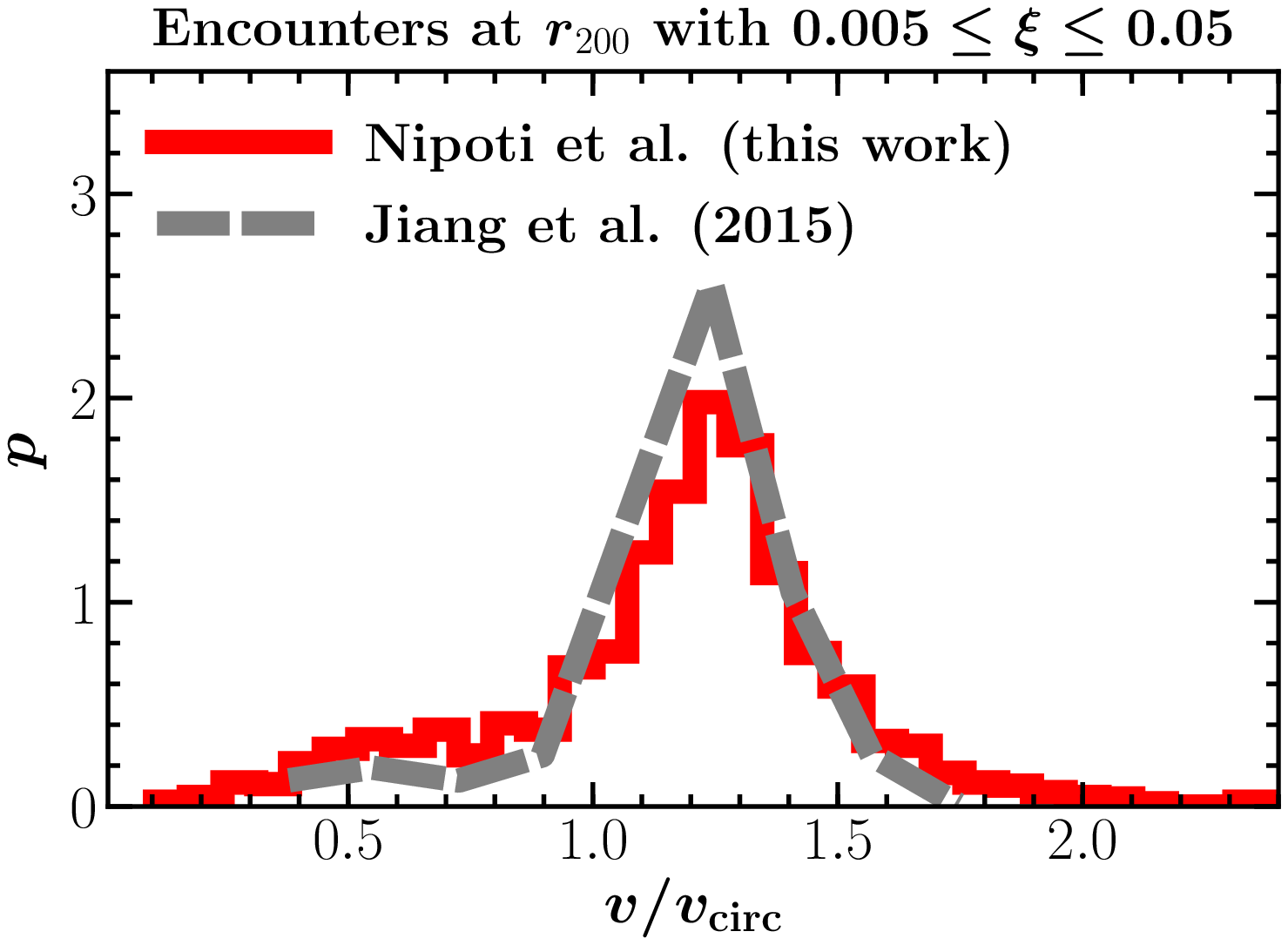,width=0.5\hsize}\psfig{file=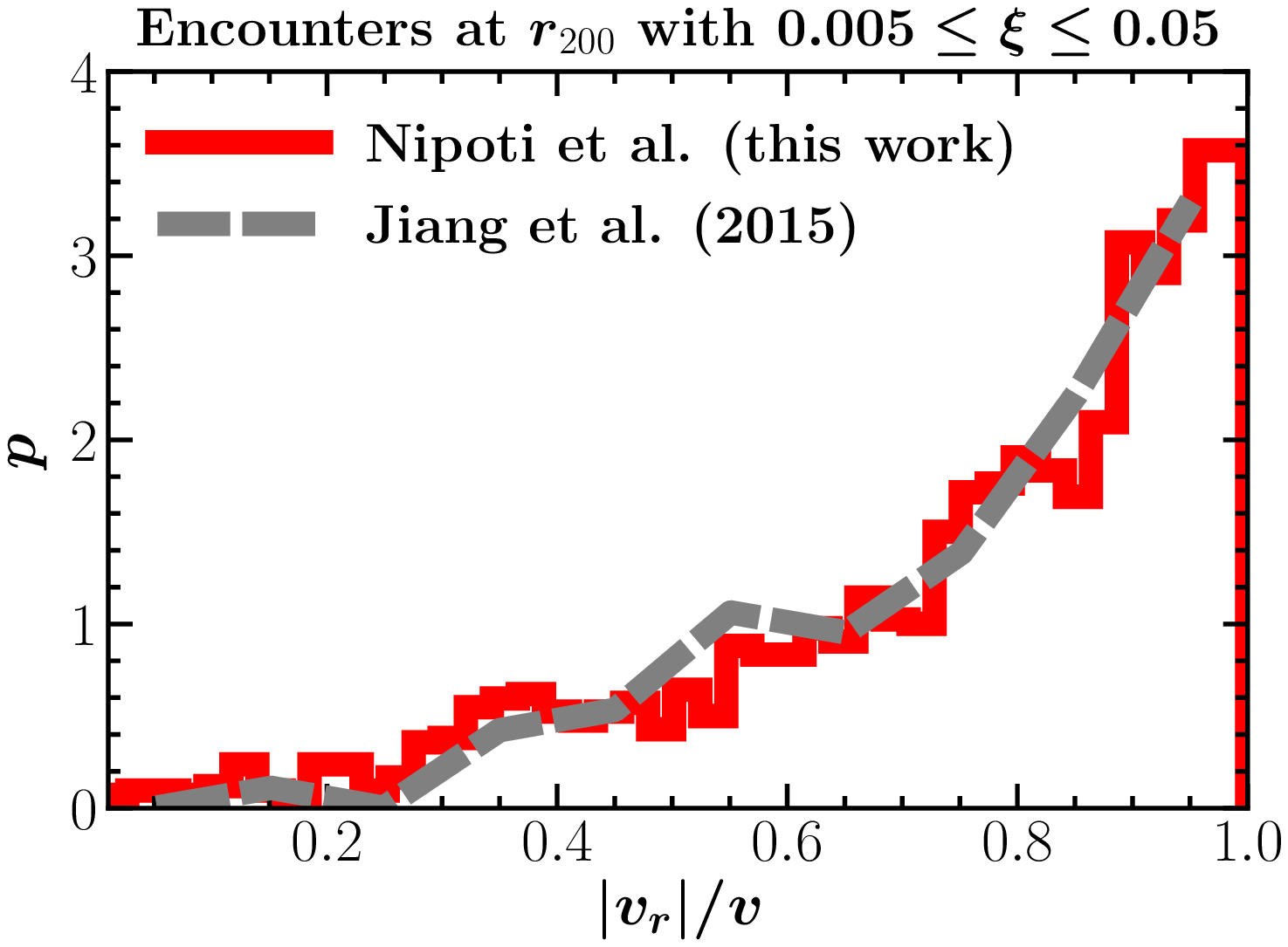,width=0.5\hsize}}
  \caption{Probability distribution $p=\d n/\d x$ of the relative
    speed ($x=v/\vcirc$; left-hand panel) and radial-to-total relative
    velocity ratio ($x=|\vr|/v$; right-hand panel) measured at
    $\rtwohc$ for halo-halo mergers in this work (solid curves) and in
    \citetalias{Jia15} (dashed curves). The solid curves are obtained
    for the cosmological simulations Ada and Bice considering mergers
    in the redshift range $0<z<1$ for haloes with $z=0$ mass
    $\Mvir\geq 10^{14}\Msun$. The dashed curves are obtained for the
    cosmological simulation DOVE considering mergers from $z=0$ up to
    the halo formation redshift (see text), for haloes with $z=0$ mass
    $\Mtwohc\approx 10^{14}\Msun$. In both cases the satellite-to-host
    mass ratio is in the range $0.005\leq \xi\leq 0.05$. }
\label{fig:vvr_200}
\end{figure*}

Here we compare the distributions of orbital parameters of the
encounters experienced by the haloes of our sample with those found by
\citetalias{Jia15} in the cosmological dark-matter only simulation
DOVE. Specifically, we consider here the distributions of $v/\vcirc$
and $|\vr|/v$ found by \citetalias{Jia15} for dark-matter haloes of
$z=0$ mass $\Mtwohc\approx 10^{14}\Msun$ considering mergers with mass
ratios in the range $0.005\leq\xi\leq 0.05$ in the redshift interval
$0<z<\zHF$, where $\zHF$ is the formation redshift of the halo (for
$\Mtwohc \approx 10^{14}\Msun$ the distribution of $\zHF$ peaks
between $z=0.5$ and $z=1$; see \citetalias{Jia15}).  In order to
compare our results with those of \citetalias{Jia15}, we built the
$\Deltac=200$ merger tree of our sample of 101 haloes with $z=0$ mass
$\Mvir\geq 10^{14}\Msun$, taking all encounters with mass ratio
$0.005\leq \xi \leq 0.05$. Altogether, in this way we select a sample
of 1855 encounters (see \Tab\ref{tab:encounters}).  For these
encounters we evaluated $v/\vcirc$ and $\vr/v$ at $\rtwohc$ using
\Eqs(\ref{eq:vcorr}) and (\ref{eq:vrcorr}). The results are shown in
\Fig\ref{fig:vvr_200}: overall the agreement between our distributions
and those of \citetalias{Jia15} is remarkable.  The peaks and the
widths of the two distributions of $v/\vcirc$ (left-hand panel of
\Fig\ref{fig:vvr_200}) almost coincide, while the tail at low values
of $v/\vcirc$ is somewhat stronger in our distribution than in the
distribution of \citetalias{Jia15}. The two distributions of $|\vr|/v$
(right-hand panel of \Fig\ref{fig:vvr_200}) are in excellent agreement
over the entire range $0\leq |\vr|/v\leq 1$. It must be noted that the
time sampling of the DOVE simulation is significantly better than that
of Le SBARBINE: for instance, in the redshift range $0\lesssim
z\lesssim 1$ the number of available snapshots is 38 for DOVE and 13
for Le SBARBINE. Thus, the agreement in the distributions of
$v/\vcirc$ and $|\vr|/v$ between \citetalias{Jia15}'s sample of
encounters and ours suggests that the correction
(\Sect\ref{sec:orbital_parameters}) we applied to estimate the
parameters at separation $\rDelta$ should be reliable.

%%%%%%%%%%%%%% FIG A1
%%%%%%%%%%%%%%%%%%%%%%%
\bibliography{hal}

\begin{thebibliography}{}
\makeatletter
\relax
\def\mn@urlcharsother{\let\do\@makeother \do\$\do\&\do\#\do\^\do\_\do\%\do\~}
\def\mn@doi{\begingroup\mn@urlcharsother \@ifnextchar [ {\mn@doi@}
  {\mn@doi@[]}}
\def\mn@doi@[#1]#2{\def\@tempa{#1}\ifx\@tempa\@empty \href
  {http://dx.doi.org/#2} {doi:#2}\else \href {http://dx.doi.org/#2} {#1}\fi
  \endgroup}
\def\mn@eprint#1#2{\mn@eprint@#1:#2::\@nil}
\def\mn@eprint@arXiv#1{\href {http://arxiv.org/abs/#1} {{\tt arXiv:#1}}}
\def\mn@eprint@dblp#1{\href {http://dblp.uni-trier.de/rec/bibtex/#1.xml}
  {dblp:#1}}
\def\mn@eprint@#1:#2:#3:#4\@nil{\def\@tempa {#1}\def\@tempb {#2}\def\@tempc
  {#3}\ifx \@tempc \@empty \let \@tempc \@tempb \let \@tempb \@tempa \fi \ifx
  \@tempb \@empty \def\@tempb {arXiv}\fi \@ifundefined
  {mn@eprint@\@tempb}{\@tempb:\@tempc}{\expandafter \expandafter \csname
  mn@eprint@\@tempb\endcsname \expandafter{\@tempc}}}

\bibitem[\protect\citeauthoryear{{Bellstedt} et~al.,}{{Bellstedt}
  et~al.}{2016}]{Bel16}
{Bellstedt} S.,  et~al., 2016, \mn@doi [\mnras] {10.1093/mnras/stw1184}, \href
  {http://adsabs.harvard.edu/abs/2016MNRAS.460.2862B} {460, 2862}

\bibitem[\protect\citeauthoryear{{Benson}}{{Benson}}{2005}]{Ben05}
{Benson} A.~J.,  2005, \mn@doi [\mnras] {10.1111/j.1365-2966.2005.08788.x},
  \href {http://adsabs.harvard.edu/abs/2005MNRAS.358..551B} {358, 551}

\bibitem[\protect\citeauthoryear{{Bernardi}, {Hyde}, {Sheth}, {Miller}  \&
  {Nichol}}{{Bernardi} et~al.}{2007}]{Ber07}
{Bernardi} M.,  {Hyde} J.~B.,  {Sheth} R.~K.,  {Miller} C.~J.,   {Nichol}
  R.~C.,  2007, \mn@doi [\aj] {10.1086/511783}, \href
  {http://adsabs.harvard.edu/abs/2007AJ....133.1741B} {133, 1741}

\bibitem[\protect\citeauthoryear{{Bernardi}, {Roche}, {Shankar}  \&
  {Sheth}}{{Bernardi} et~al.}{2011}]{Ber11}
{Bernardi} M.,  {Roche} N.,  {Shankar} F.,   {Sheth} R.~K.,  2011, \mn@doi
  [\mnras] {10.1111/j.1365-2966.2010.17984.x}, \href
  {http://adsabs.harvard.edu/abs/2011MNRAS.412..684B} {412, 684}

\bibitem[\protect\citeauthoryear{{Binney} \& {Tremaine}}{{Binney} \&
  {Tremaine}}{1987}]{Bin87}
{Binney} J.,  {Tremaine} S.,  1987, {Galactic dynamics}.
Princeton University Press, Princeton, NJ

\bibitem[\protect\citeauthoryear{{Boylan-Kolchin}, {Ma}  \&
  {Quataert}}{{Boylan-Kolchin} et~al.}{2006}]{Boy06}
{Boylan-Kolchin} M.,  {Ma} C.-P.,   {Quataert} E.,  2006, \mn@doi [\mnras]
  {10.1111/j.1365-2966.2006.10379.x}, \href
  {http://adsabs.harvard.edu/abs/2006MNRAS.369.1081B} {369, 1081}

\bibitem[\protect\citeauthoryear{{Buchan} \& {Shankar}}{{Buchan} \&
  {Shankar}}{2016}]{Buc16}
{Buchan} S.,  {Shankar} F.,  2016, \mn@doi [\mnras] {10.1093/mnras/stw1771},
  \href {http://adsabs.harvard.edu/abs/2016MNRAS.462.2001B} {462, 2001}

\bibitem[\protect\citeauthoryear{{Chandrasekhar}}{{Chandrasekhar}}{1943}]{Cha43}
{Chandrasekhar} S.,  1943, \mn@doi [\apj] {10.1086/144517}, \href
  {http://adsabs.harvard.edu/abs/1943ApJ....97..255C} {97, 255}

\bibitem[\protect\citeauthoryear{{Chua}, {Pillepich}, {Rodriguez-Gomez},
  {Vogelsberger}, {Bird}  \& {Hernquist}}{{Chua} et~al.}{2017}]{Chu17}
{Chua} K.~T.~E.,  {Pillepich} A.,  {Rodriguez-Gomez} V.,  {Vogelsberger} M.,
  {Bird} S.,   {Hernquist} L.,  2017, \mn@doi [\mnras] {10.1093/mnras/stx2238},
  \href {http://adsabs.harvard.edu/abs/2017MNRAS.472.4343C} {472, 4343}

\bibitem[\protect\citeauthoryear{{De Lucia} \& {Blaizot}}{{De Lucia} \&
  {Blaizot}}{2007}]{Del07}
{De Lucia} G.,  {Blaizot} J.,  2007, \mn@doi [\mnras]
  {10.1111/j.1365-2966.2006.11287.x}, \href
  {http://adsabs.harvard.edu/abs/2007MNRAS.375....2D} {375, 2}

\bibitem[\protect\citeauthoryear{{Despali}, {Giocoli}, {Angulo}, {Tormen},
  {Sheth}, {Baso}  \& {Moscardini}}{{Despali} et~al.}{2016}]{Des16}
{Despali} G.,  {Giocoli} C.,  {Angulo} R.~E.,  {Tormen} G.,  {Sheth} R.~K.,
  {Baso} G.,   {Moscardini} L.,  2016, \mn@doi [\mnras]
  {10.1093/mnras/stv2842}, \href
  {http://adsabs.harvard.edu/abs/2016MNRAS.456.2486D} {456, 2486}

\bibitem[\protect\citeauthoryear{{Despali}, {Giocoli}, {Bonamigo}, {Limousin}
  \& {Tormen}}{{Despali} et~al.}{2017}]{Des17}
{Despali} G.,  {Giocoli} C.,  {Bonamigo} M.,  {Limousin} M.,   {Tormen} G.,
  2017, \mn@doi [\mnras] {10.1093/mnras/stw3129}, \href
  {http://adsabs.harvard.edu/abs/2017MNRAS.466..181D} {466, 181}

\bibitem[\protect\citeauthoryear{{Dubinski}}{{Dubinski}}{1998}]{Dub98}
{Dubinski} J.,  1998, \mn@doi [\apj] {10.1086/305901}, \href
  {http://adsabs.harvard.edu/abs/1998ApJ...502..141D} {502, 141}

\bibitem[\protect\citeauthoryear{{Eke}, {Cole}  \& {Frenk}}{{Eke}
  et~al.}{1996}]{Eke96}
{Eke} V.~R.,  {Cole} S.,   {Frenk} C.~S.,  1996, \mn@doi [\mnras]
  {10.1093/mnras/282.1.263}, \href
  {http://adsabs.harvard.edu/abs/1996MNRAS.282..263E} {282}

\bibitem[\protect\citeauthoryear{{Feldmann}, {Carollo}, {Mayer}, {Renzini},
  {Lake}, {Quinn}, {Stinson}  \& {Yepes}}{{Feldmann} et~al.}{2010}]{Fel10}
{Feldmann} R.,  {Carollo} C.~M.,  {Mayer} L.,  {Renzini} A.,  {Lake} G.,
  {Quinn} T.,  {Stinson} G.~S.,   {Yepes} G.,  2010, \mn@doi [\apj]
  {10.1088/0004-637X/709/1/218}, \href
  {http://adsabs.harvard.edu/abs/2010ApJ...709..218F} {709, 218}

\bibitem[\protect\citeauthoryear{{Giocoli}}{{Giocoli}}{2008}]{giocoli08thesis}
{Giocoli} C.,  2008, Hierarchical Clustering: Structure Formation in the
  Universe.
Phd Thesis, University of Padova, Padova, Italy,
  http://paduaresearch.cab.unipd.it/850/

\bibitem[\protect\citeauthoryear{{Giocoli}}{{Giocoli}}{2010}]{giocoli10c}
{Giocoli} C.,  2010, in {J.-M.~Alimi \& A.~Fu{\"o}zfa} ed.,  American Institute
  of Physics Conference Series Vol. 1241, American Institute of Physics
  Conference Series. pp 892--897 (\mn@eprint {arXiv} {0911.2969}),
  \mn@doi{10.1063/1.3462730}

\bibitem[\protect\citeauthoryear{{Giocoli}, {Tormen}  \& {van den
  Bosch}}{{Giocoli} et~al.}{2008}]{giocoli08}
{Giocoli} C.,  {Tormen} G.,   {van den Bosch} F.~C.,  2008, \mn@doi [\mnras]
  {10.1111/j.1365-2966.2008.13182.x}, \href
  {http://adsabs.harvard.edu/abs/2008MNRAS.386.2135G} {386, 2135}

\bibitem[\protect\citeauthoryear{{Giocoli}, {Tormen}  \& {Sheth}}{{Giocoli}
  et~al.}{2012}]{giocoli12b}
{Giocoli} C.,  {Tormen} G.,   {Sheth} R.~K.,  2012, \mn@doi [\mnras]
  {10.1111/j.1365-2966.2012.20594.x}, \href
  {http://adsabs.harvard.edu/abs/2012MNRAS.422..185G} {422, 185}

\bibitem[\protect\citeauthoryear{{Hausman} \& {Ostriker}}{{Hausman} \&
  {Ostriker}}{1978}]{Hau78}
{Hausman} M.~A.,  {Ostriker} J.~P.,  1978, \mn@doi [\apj] {10.1086/156380},
  \href {http://adsabs.harvard.edu/abs/1978ApJ...224..320H} {224, 320}

\bibitem[\protect\citeauthoryear{{Jiang}, {Cole}, {Sawala}  \& {Frenk}}{{Jiang}
  et~al.}{2015}]{Jia15}
{Jiang} L.,  {Cole} S.,  {Sawala} T.,   {Frenk} C.~S.,  2015, \mn@doi [\mnras]
  {10.1093/mnras/stv053}, \href
  {http://adsabs.harvard.edu/abs/2015MNRAS.448.1674J} {448, 1674}

\bibitem[\protect\citeauthoryear{{Khochfar} \& {Burkert}}{{Khochfar} \&
  {Burkert}}{2006}]{Kho06}
{Khochfar} S.,  {Burkert} A.,  2006, \mn@doi [\aap]
  {10.1051/0004-6361:20053241}, \href
  {http://adsabs.harvard.edu/abs/2006A%26A...445..403K} {445, 403}

\bibitem[\protect\citeauthoryear{{Kravtsov}}{{Kravtsov}}{2013}]{Kra13}
{Kravtsov} A.~V.,  2013, \mn@doi [\apjl] {10.1088/2041-8205/764/2/L31}, \href
  {http://adsabs.harvard.edu/abs/2013ApJ...764L..31K} {764, L31}

\bibitem[\protect\citeauthoryear{{Lauer}, {Postman}, {Strauss}, {Graves}  \&
  {Chisari}}{{Lauer} et~al.}{2014}]{Lau14}
{Lauer} T.~R.,  {Postman} M.,  {Strauss} M.~A.,  {Graves} G.~J.,   {Chisari}
  N.~E.,  2014, \mn@doi [\apj] {10.1088/0004-637X/797/2/82}, \href
  {http://adsabs.harvard.edu/abs/2014ApJ...797...82L} {797, 82}

\bibitem[\protect\citeauthoryear{{Lidman} et~al.,}{{Lidman}
  et~al.}{2013}]{Lid13}
{Lidman} C.,  et~al., 2013, \mn@doi [\mnras] {10.1093/mnras/stt777}, \href
  {http://adsabs.harvard.edu/abs/2013MNRAS.433..825L} {433, 825}

\bibitem[\protect\citeauthoryear{{Liu}, {Xia}, {Mao}, {Wu}  \& {Deng}}{{Liu}
  et~al.}{2008}]{Liu08}
{Liu} F.~S.,  {Xia} X.~Y.,  {Mao} S.,  {Wu} H.,   {Deng} Z.~G.,  2008, \mn@doi
  [\mnras] {10.1111/j.1365-2966.2007.12818.x}, \href
  {http://adsabs.harvard.edu/abs/2008MNRAS.385...23L} {385, 23}

\bibitem[\protect\citeauthoryear{{Lynden-Bell}}{{Lynden-Bell}}{1967}]{Lyn67}
{Lynden-Bell} D.,  1967, \mn@doi [\mnras] {10.1093/mnras/136.1.101}, \href
  {http://adsabs.harvard.edu/abs/1967MNRAS.136..101L} {136, 101}

\bibitem[\protect\citeauthoryear{{Marchesini} et~al.,}{{Marchesini}
  et~al.}{2014}]{Mar14}
{Marchesini} D.,  et~al., 2014, \mn@doi [\apj] {10.1088/0004-637X/794/1/65},
  \href {http://adsabs.harvard.edu/abs/2014ApJ...794...65M} {794, 65}

\bibitem[\protect\citeauthoryear{{Merritt}}{{Merritt}}{1985}]{Mer85}
{Merritt} D.,  1985, \mn@doi [\apj] {10.1086/162860}, \href
  {http://adsabs.harvard.edu/abs/1985ApJ...289...18M} {289, 18}

\bibitem[\protect\citeauthoryear{{Naab}, {Johansson}  \& {Ostriker}}{{Naab}
  et~al.}{2009}]{Naa09}
{Naab} T.,  {Johansson} P.~H.,   {Ostriker} J.~P.,  2009, \mn@doi [\apjl]
  {10.1088/0004-637X/699/2/L178}, \href
  {http://adsabs.harvard.edu/abs/2009ApJ...699L.178N} {699, L178}

\bibitem[\protect\citeauthoryear{{Nipoti}}{{Nipoti}}{2017}]{Nip17}
{Nipoti} C.,  2017, \mn@doi [\mnras] {10.1093/mnras/stx112}, \href
  {http://adsabs.harvard.edu/abs/2017MNRAS.467..661N} {467, 661}

\bibitem[\protect\citeauthoryear{{Nipoti}, {Treu}  \& {Bolton}}{{Nipoti}
  et~al.}{2009}]{Nip09}
{Nipoti} C.,  {Treu} T.,   {Bolton} A.~S.,  2009, \mn@doi [\apj]
  {10.1088/0004-637X/703/2/1531}, \href
  {http://adsabs.harvard.edu/abs/2009ApJ...703.1531N} {703, 1531}

\bibitem[\protect\citeauthoryear{{Nipoti}, {Treu}, {Leauthaud}, {Bundy},
  {Newman}  \& {Auger}}{{Nipoti} et~al.}{2012}]{Nip12}
{Nipoti} C.,  {Treu} T.,  {Leauthaud} A.,  {Bundy} K.,  {Newman} A.~B.,
  {Auger} M.~W.,  2012, \mn@doi [\mnras] {10.1111/j.1365-2966.2012.20749.x},
  \href {http://adsabs.harvard.edu/abs/2012MNRAS.422.1714N} {422, 1714}

\bibitem[\protect\citeauthoryear{{Ostriker} \& {Tremaine}}{{Ostriker} \&
  {Tremaine}}{1975}]{Ost75}
{Ostriker} J.~P.,  {Tremaine} S.~D.,  1975, \mn@doi [\apjl] {10.1086/181992},
  \href {http://adsabs.harvard.edu/abs/1975ApJ...202L.113O} {202, L113}

\bibitem[\protect\citeauthoryear{{Planck Collaboration} et~al.,}{{Planck
  Collaboration} et~al.}{2014}]{Pla14}
{Planck Collaboration} et~al., 2014, \mn@doi [\aap]
  {10.1051/0004-6361/201321591}, \href
  {http://adsabs.harvard.edu/abs/2014A%26A...571A..16P} {571, A16}

\bibitem[\protect\citeauthoryear{{Posti}, {Nipoti}, {Stiavelli}  \&
  {Ciotti}}{{Posti} et~al.}{2014}]{Pos14}
{Posti} L.,  {Nipoti} C.,  {Stiavelli} M.,   {Ciotti} L.,  2014, \mn@doi
  [\mnras] {10.1093/mnras/stu301}, \href
  {http://adsabs.harvard.edu/abs/2014MNRAS.440..610P} {440, 610}

\bibitem[\protect\citeauthoryear{{Rodriguez-Gomez} et~al.,}{{Rodriguez-Gomez}
  et~al.}{2016}]{Rod16}
{Rodriguez-Gomez} V.,  et~al., 2016, \mn@doi [\mnras] {10.1093/mnras/stw456},
  \href {http://adsabs.harvard.edu/abs/2016MNRAS.458.2371R} {458, 2371}

\bibitem[\protect\citeauthoryear{{Ruszkowski} \& {Springel}}{{Ruszkowski} \&
  {Springel}}{2009}]{Rus09}
{Ruszkowski} M.,  {Springel} V.,  2009, \mn@doi [\apj]
  {10.1088/0004-637X/696/2/1094}, \href
  {http://adsabs.harvard.edu/abs/2009ApJ...696.1094R} {696, 1094}

\bibitem[\protect\citeauthoryear{{Shankar} et~al.,}{{Shankar}
  et~al.}{2015}]{Sha15}
{Shankar} F.,  et~al., 2015, \mn@doi [\apj] {10.1088/0004-637X/802/2/73}, \href
  {http://adsabs.harvard.edu/abs/2015ApJ...802...73S} {802, 73}

\bibitem[\protect\citeauthoryear{{Tonini}, {Bernyk}, {Croton}, {Maraston}  \&
  {Thomas}}{{Tonini} et~al.}{2012}]{Ton12}
{Tonini} C.,  {Bernyk} M.,  {Croton} D.,  {Maraston} C.,   {Thomas} D.,  2012,
  \mn@doi [\apj] {10.1088/0004-637X/759/1/43}, \href
  {http://adsabs.harvard.edu/abs/2012ApJ...759...43T} {759, 43}

\bibitem[\protect\citeauthoryear{{Tormen}}{{Tormen}}{1998}]{tormen98}
{Tormen} G.,  1998, \mnras, \href
  {http://adsabs.harvard.edu/abs/1998MNRAS.297..648T} {297, 648}

\bibitem[\protect\citeauthoryear{{Tormen}, {Moscardini}  \& {Yoshida}}{{Tormen}
  et~al.}{2004}]{tormen04}
{Tormen} G.,  {Moscardini} L.,   {Yoshida} N.,  2004, \mn@doi [\mnras]
  {10.1111/j.1365-2966.2004.07736.x}, \href
  {http://adsabs.harvard.edu/abs/2004MNRAS.350.1397T} {350, 1397}

\bibitem[\protect\citeauthoryear{{Tremaine}}{{Tremaine}}{1990}]{Tre90}
{Tremaine} S.,  1990, {The origin of central cluster galaxies.}.
pp 394--405

\bibitem[\protect\citeauthoryear{{Volonteri} \& {Ciotti}}{{Volonteri} \&
  {Ciotti}}{2013}]{Vol13}
{Volonteri} M.,  {Ciotti} L.,  2013, \mn@doi [\apj]
  {10.1088/0004-637X/768/1/29}, \href
  {http://adsabs.harvard.edu/abs/2013ApJ...768...29V} {768, 29}

\bibitem[\protect\citeauthoryear{{Vulcani} et~al.,}{{Vulcani}
  et~al.}{2014}]{Vul14}
{Vulcani} B.,  et~al., 2014, \mn@doi [\apj] {10.1088/0004-637X/797/1/62}, \href
  {http://adsabs.harvard.edu/abs/2014ApJ...797...62V} {797, 62}

\bibitem[\protect\citeauthoryear{{Vulcani} et~al.,}{{Vulcani}
  et~al.}{2016}]{Vul16b}
{Vulcani} B.,  et~al., 2016, \mn@doi [\apj] {10.3847/0004-637X/816/2/86}, \href
  {http://adsabs.harvard.edu/abs/2016ApJ...816...86V} {816, 86}

\bibitem[\protect\citeauthoryear{{White}}{{White}}{1976}]{Whi76}
{White} S.~D.~M.,  1976, \mn@doi [\mnras] {10.1093/mnras/174.1.19}, \href
  {http://adsabs.harvard.edu/abs/1976MNRAS.174...19W} {174, 19}

\makeatother
\end{thebibliography}
\bibliographystyle{mnras}
\end{document}